\documentclass[]{spie}  %>>> use for US letter paper
\usepackage{amsmath, amssymb, amsfonts, amscd, mathrsfs, color, relsize, graphicx}
\usepackage[colorlinks=true, allcolors=blue]{hyperref}
\usepackage{listings}
\allowdisplaybreaks

\newtheorem{theo}{Theorem}[section]
\newtheorem{defi}[theo]{Definition}
\newtheorem{lem}[theo]{Lemma}

\newtheorem{cor}[theo]{Corollary}
\newtheorem{rem}[theo]{Remark}

\numberwithin{equation}{section}

\newcommand{\R}{\mathbb{R}} % reals
\newcommand{\C}{\mathbb{C}} % complexes
\newcommand{\Z}{\mathbb{Z}} % integers
\newcommand{\N}{\mathbb{N}} % naturals
\newcommand{\G}{\Gamma} % group

\newcommand{\T}{\mathcal{T}_{{}_\G}} % periodization mapping
\newcommand{\pj}{\mathcal{P}}        % projection for the range function

\newcommand{\Proj}{\mathbb{P}} % projections

\newcommand{\vsp}{\textnormal{span}}
\newcommand{\ol}[1]{\overline{#1}}

\newcommand{\wh}[1]{\widehat{#1}}

\newcommand{\data}{\mathscr{F}}        % data
\newcommand{\rdata}{\mathcal{X}}       % augmented fibered data
\newcommand{\D}{\mathscr{E}}           % reduced error functional
\newcommand{\W}{\mathcal{W}}           % fiber subspace
\newcommand{\V}{\mathcal{V}}           % minimal fiber subspace

\title{Optimal translational-rotational invariant dictionaries\\ for images}

\author[a]{Davide Barbieri}
\author[b]{Carlos Cabrelli}
\author[c]{Eugenio Hern\'andez}
\author[d]{Ursula Molter}
\affil[a, c]{Departamento de Matem\'aticas, Universidad Aut\'onoma de Madrid, 28049, Madrid, Spain}
\affil[b, d]{Departamento de Matem\'atica, Universidad de Buenos Aires, and Instituto de Matem\'atica ``Luis Santal\'o'' (IMAS-CONICET-UBA), 1428 Buenos Aires, Argentina}

\authorinfo{Further author information\\ E-mail: davide.barbieri@uam.es}

\begin{document} 
\maketitle

\begin{abstract}
We provide the construction of a set of square matrices whose translates and rotates provide a Parseval frame that is optimal for approximating a given dataset of images. Our approach is based on abstract harmonic analysis techniques. Optimality is considered with respect to the quadratic error of approximation of the images in the dataset with their projection onto a linear subspace that is invariant under translations and rotations. In addition, we provide an elementary and fully self-contained proof of optimality, and the numerical results from datasets of natural images.
\end{abstract}


\section{Introduction} \label{sec:intro} 

The purpose of this work is to present the theoretical solution, and outline the numerical implementation, for an optimal approximation problem on digital images. The problem is the following: given a dataset of square images, we want to find the optimal generators that provide, by translations and 90 degrees rotations, the best approximation of the dataset with respect to a quadratic error.

We are not considering the full set of all possible translations, which would give rise to a convolutional problem, but rather we consider translations on a lattice. Together with rotations, they will define a nonabelian semidirect product group of discrete Euclidean rigid movements of images.

This work is an adaptation of a result obtained for more general groups in [\citenum{BCHM2019}], which nevertheless can not be directly applied to this setting. Our approach to invariant approximation borrows several ideas from the theory of approximation by shift-invariant spaces developed in [\citenum{ACHM2007}], see also [\citenum{AT2011, CMP2017}]. Noncommutative problems of harmonic analysis related to group actions have a long tradition in signal processing, and recent works with relevant interactions with the present one are [\citenum{BHP2015, BHP2018, GHP2018, GHI2018}].

The presence of invariances in natural images has been long studied and exploited in vision (see [\citenum{CS2006, B2015, PA2016}] and references therein), and it plays a central role in several approaches to machine learning [\citenum{Bekkers2018, AERP2019}]. In particular, the solution presented in this work makes use of a special form of a data augmentation, a classical technique now of common use for networks training (see e.g. [\citenum{DM2001, google2019}] and references therein). Our approach differs in a fundamental way from patch-based ones such as [\citenum{OF1996, AEB2006}], because we do not extract patches from images, to be then used by translations, but rather consider entire images, and find the optimal generators for a fixed set of translations and rotations with methods of Fourier analysis.

The structure of this paper is the following. In Section \ref{sec:preliminaries} we describe the group invariance, focusing on invariant subspaces, and provide a formal statement of the approximation problem. In Section \ref{sec:Fourier} we introduce an isometric isomophism that allows us to treat the group symmetries with Fourier analysis, and study how invariant subspaces behave under such a map. In Section \ref{sec:solution} we provide a formal statement of the proposed solution, and outline the algorithm that allows us to compute it. In Section \ref{sec:numerics} we finally show the numerical results on a well-known dataset of natural images.

Most of the theoretical results presented in this paper could be deduced, without major difficulties, from the ones obtained in [\citenum{BCHM2019}]. The only obstruction from applying them directly to the present setting is due to the fact that the hypotheses for Proposition 4.1 of [\citenum{BCHM2019}] are not met here due to the presence of certain nontrivial stabilizers for the group action. In Section \ref{sec:Fourier} we overcome this issue by defining an isometry that is slightly different from the one introduced in Section 4.2 of [\citenum{BCHM2019}]. The present setting represents a great simplification of the general case, mainly due to the finiteness of the problem. This gives us the possibility to present a fully self-contained approach to the solution. Indeed, although the arguments used to solve this problem refer to much more general principles, and could be proved with more abstract techniques, in the present case it is possible to provide full proofs of all the results needed to construct the desired approximation with only elementary techniques. We have chosen to do so with the intention of making this work accessible to the non specialist reader, which may be interested in applying this technique.

\section{The invariant approximation problem}\label{sec:preliminaries}

\subsection{Group Invariance}

We will consider grayscale digital images of $d \times d$ pixels, and we will treat here the case of an odd number $d$. It is convenient, for the purposes of this work, to consider a digital image $f$ as a function on the square lattice in \emph{centered coordinates}
$$
\Z_d \times \Z_d = \left\{n = (n_1,n_2) : n_1, n_2 \in \left\{-\frac{d-1}{2},\dots,\frac{d-1}{2}\right\}\right\} ,
$$
i.e. $f \in \ell_2(\Z_d \times \Z_d)$. This space is $\C^{d^2}$, indexed by $\Z_d \times \Z_d$, and endowed with the Euclidean norm, that we denote by $\|\cdot\|_d$, associated to the inner product
\begin{equation}\label{eq:norm}
\langle f, f'\rangle_d = \sum_{n_1 = -\frac{d-1}{2}}^{\frac{d-1}{2}} \sum_{n_2 = -\frac{d-1}{2}}^{\frac{d-1}{2}} f(n_1,n_2) \ol{f'(n_1,n_2)} , \quad f, f' \in \ell_2(\Z_d \times \Z_d).
\end{equation}
In particular, $\|f\|_d$ is the Frobenius norm of $f$ viewed as a matrix.

Note that, for simplicity, we allow ourselves the slight abuse of keeping the same notation $\Z_d \times \Z_d$ commonly used for $\{0,\dots,d\} \times \{0,\dots,d\}$. For any $n, n' \in \Z_d \times \Z_d$ we will also keep the additive notation, and denote always by $n + n' \in \Z_d \times \Z_d$ the periodic sum
\begin{equation}\label{eq:composition}
n + n' = \Big( (n_1 + n_1' + \frac{d-1}{2}) \!\!\!\! \mod d \ - \frac{d-1}{2} \, , \, (n_2 + n_2' + \frac{d-1}{2}) \!\!\!\! \mod d \ - \frac{d-1}{2} \Big).
\end{equation}
With this operation, $\Z_d \times \Z_d$ is an abelian group, and 90 degrees rotations, defined by the linear action on $\Z_d \times \Z_d$ of
\begin{equation}\label{eq:rotation}
r = \left(\begin{array}{cc}
0 & -1\\ 1 & 0
\end{array}\right) ,
\end{equation}
are automorphisms. This can be easily checked because $r(n_1,n_2) = (-n_2,n_1) \in \Z_d \times \Z_d$ and, using (\ref{eq:composition}), $r(n + n') = r(n) + r(n')$. For $g \in \{0,1,2,3\}$ we denote by $r^g$ the $g$-th power of the matrix $r$, so $r^2$ corresponds to a 180 degrees rotation, etc. Note that $r^4$ is the identity: $G = \{r^g\}_{g = 0}^3$ is a cyclic group of order 4.
% Proof that r is actually an automorphism:
% \begin{align*}
% r(n + n') & = \Big( - (n_2 + n_2' + \frac{d-1}{2}) \!\!\!\! \mod d \ + \frac{d-1}{2} \, , \, (n_1 + n_1' + \frac{d-1}{2}) \!\!\!\! \mod d \ - \frac{d-1}{2} \Big)\\
% r(n) + r(n') & = \Big( (- n_2 - n_2' + \frac{d-1}{2}) \!\!\!\! \mod d \ - \frac{d-1}{2} \, , \, (n_1 + n_1' + \frac{d-1}{2}) \!\!\!\! \mod d \ - \frac{d-1}{2} \Big)
% \end{align*}

Moreover, whenever $d$ is not a prime number, $\Z_d \times \Z_d$ admits nontrivial proper subgroups for which 90 degrees rotations are also automorphisms of these subgroups. We give a precise statement in the next lemma.

\begin{lem}\label{lem:subgroups}
Let $p, q$ be odd, and let $d = pq$.
\begin{itemize}
\item[i)] $\Lambda = \Big\{\lambda = (t_1,t_2)q : t_1, t_2 \in \big\{-\frac{p-1}{2},\dots,\frac{p-1}{2}\big\}\Big\} \subset \Z_d \times \Z_d$ is a subgroup, isomorphic to $\Z_p \times \Z_p$.
\item[ii)] $\Lambda$ is invariant for $r$, i.e. $r(\Lambda) = \Lambda$, and for $\lambda, \lambda' \in \Lambda$ we have $r(\lambda + \lambda') = r(\lambda) + r(\lambda')$.
\end{itemize}
\end{lem}

\newpage
On images, and on every $f \in \ell_2(\Z_d \times \Z_d)$, the symmetries of translations and rotations are formally described as follows. Let $\Lambda \subset \Z_d \times \Z_d$ be a subgroup. For $\lambda = (\lambda_1, \lambda_2) \in \Lambda$, the translation of $f$ by $\lambda$, denoted by $T(\lambda)f \in \ell_2(\Z_d \times \Z_d)$, is
$$
T(\lambda)f(n) = f(n-\lambda) \quad n \in \Z_d \times \Z_d
$$
where the operation $n - \lambda$ is intended as in (\ref{eq:composition}). The 90 degrees rotation of $f$, denoted by $Rf \in \ell_2(\Z_d \times \Z_d)$, is
$$
Rf(n_1,n_2) = f\big(r(n_1,n_2)\big) = f(-n_2,n_1)
$$
where $r$ is the 90 degrees rotation of $\Z_d \times \Z_d$ given by (\ref{eq:rotation}). For $g \in \{0,1,2,3\}$ we denote by $R^g$ the $g$-th power (iteration) of the operator $R$. The set of operators $\{T(\lambda)R^g : \lambda \in \Lambda, g \in \{0,1,2,3\}\}$ define a unitary representation of the nonabelian \emph{group} $\G = \Lambda \rtimes G$ of discrete Euclidean rigid movements, i.e. $\Lambda$-translations and 90 degrees rotations on $\Z_d \times \Z_d$. The composition law of $\G$ can be written as
$$
(\lambda,g)\cdot(\lambda,g') = (\lambda + r^{-g}(\lambda'),g + g') , \quad 
$$
for $\lambda, \lambda' \in \Lambda$ and $g,g' \in \{0,1,2,3\}$. Indeed, 
\begin{equation}\label{eq:commutator}
T(\lambda)R^gT(\lambda')R^{g'}f(n) = f\big(r^{g'}(r^g(n - \lambda) - \lambda')\big) = T(\lambda + r^{-g}(\lambda'))R^{g + g'}f(n) , \quad n \in \Z_d \times \Z_d ,
\end{equation}
and, for all $\lambda \in \Lambda$, $g \in \{0,1,2,3\}$ and all $f \in \ell_2(\Z_d \times \Z_d)$ we have $\|T(\lambda)R^gf\|_d = \|f\|_d$.
% $$
% \langle T(\lambda)R^g f, f'\rangle_d = \langle f, R^{-g} T(-\lambda)f'\rangle_d = \langle f,  T(-r^g(\lambda))R^{-g}f'\rangle_d, \quad \forall \ f, f' \in \ell_2(\Z_d \times \Z_d) .
% $$
Note that we are always using the composition (\ref{eq:composition}) for $\Z_d \times \Z_d$ variables, while periodic composition is used for the rotation variables, i.e. $g + g'$ is considered mod 4.

By general arguments, see \cite[Lemma 11]{BHP2018}, a subspace $V \subset \ell_2(\Z_d\times\Z_d)$ is invariant under the action of the group $\G = \Lambda \rtimes G$, i.e. it is such that $f \in V \Rightarrow T(\lambda)R^g f \in V$ for all $\lambda \in \Lambda, g \in \{0,1,2,3\}$, if and only if it is linearly generated by the $\G$-orbit of a set of vectors of $\ell_2(\Z_d\times\Z_d)$. This is the object of the next definition.
\begin{defi}\label{def:GSIS}
For a set $\Psi = \{\psi_j\}_{j=1}^\kappa \subset \ell_2(\Z_d\times\Z_d)$ of $\kappa$ \emph{generators}, we denote by
$$
S(\Psi) = \vsp\Big\{T(\lambda)R^g\psi_j \, : \, \lambda \in \Lambda, g \in \{0,1,2,3\}, j \in \{1,\dots,\kappa\}\Big\}
$$
the $\G$-invariant linear subspace of $\ell_2(\Z_d\times\Z_d)$ generated by the action of $\G$ on $\Psi$.
\end{defi}

\subsection{Best Approximation}\label{sec:problem}

The best approximation problem solved in this paper is the following. Suppose we are given a dataset $\data = \{f_1,\dots,f_m\}$ of $d \times d$ digital images. For $\kappa \in \N$, we want to find a family $\Phi = \{\phi_j\}_{j=1}^\kappa \subset \ell_2(\Z_d \times \Z_d)$ such that $\Big\{T(\lambda)R^g\phi_j \, : \, \lambda \in \Lambda, g \in \{0,1,2,3\}, j \in \{1,\dots,\kappa\}\Big\}$ is a Parseval frame of $S(\Phi)$, i.e. such that the orthogonal projection of $\ell_2(\Z_d \times \Z_d)$ onto $S(\Phi)$ can be written as
$$
\Proj_{S(\Phi)}f = \sum_{\lambda \in \Lambda} \sum_{g = 0}^3 \sum_{j= 1}^\kappa \langle f, T(\lambda)R^g\phi_j \rangle_d \, T(\lambda)R^g\phi_j , \quad f \in \ell_2(\Z_d \times \Z_d) .
$$
Moreover, we want this projection to minimize the quadratic error resulting from the projection onto a $\Gamma$-invariant space with $\kappa$ generators, i.e.
\begin{equation}\label{eq:error}
\sum_{i=1}^m \|f_i - \Proj_{S(\Phi)}f_i\|_d^2 = \min\left\{\sum_{i=1}^m \|f_i - \Proj_{S(\Psi)}f_i\|_d^2 : \Psi = \{\psi_j\}_{j=1}^\kappa \subset \ell_2(\Z_d\times\Z_d)\right\},
\end{equation}
where $\Proj_{S(\Psi)}$ is the orthogonal projection of $\ell_2(\Z_d \times \Z_d)$ onto $S(\Psi)$.

The solution to this problem will be provided in Section \ref{sec:solution} together with the construction that allows us to compute an optimal set $\Phi$ of Parseval frame generators, given in Table 1.

Observe that the dimension of a $\G$-invariant space $S(\Psi) \subset \ell_2(\Z_d\times\Z_d)$ with $\kappa$ generators is at most
$
\textnormal{dim}(S(\Psi)) \leq 4\kappa\, \sharp \Lambda.
$
If $d = pq$ and $\Lambda \approx \Z_p \times \Z_p$ is as in Lemma \ref{lem:subgroups}, then $\sharp \Lambda = p^2$. Thus, we will always consider families of generators $\Psi$ with cardinality $\kappa < \frac14 q^2.$
Indeed, this implies that the dimension of $S(\Psi)$ is certainly smaller than $d^2 = p^2q^2$, that is the dimension of $\ell_2(\Z_d \times \Z_d)$, and the ratio $\displaystyle\frac{4\kappa}{q^2}$ approximately quantifies the dimensionality reduction that one obtains when replacing $f \in \ell_2(\Z_d \times \Z_d)$ with $\Proj_{S(\Psi)}f \in S(\Psi)$.

\section{Fourier analysis of invariance}\label{sec:Fourier}

Fourier duality on $\Z_d \times \Z_d$ is provided by the DFT:
$$
\wh{f}(k_1,k_2) = \sum_{n_1 = - \frac{d-1}{2}}^{\frac{d-1}{2}} \sum_{n_2 = - \frac{d-1}{2}}^{\frac{d-1}{2}} f(n_1,n_2) e^{-2\pi i \frac{n_1k_1 + n_2k_2}{d}} , \quad k_1, k_2 \in \left\{-\frac{d-1}{2},\dots,\frac{d-1}{2}\right\}.
$$
We will write, for short,
$$
\wh{f}(k) = \sum_{n \in \Z_d \times \Z_d} f(n) e^{-2\pi i \frac{n.k}{d}} , \quad k \in \Z_d \times \Z_d
$$
where $n.k = n_1k_1 + n_2k_2$. The DFT is a multiple of an invertible isometry: it satisfies
$$\|\wh{f}\|_d = d \|f\|_d,$$ and its inverse reads
$$
f(n) = \frac{1}{d^2} \sum_{k \in \Z_d \times \Z_d} \wh{f}(k) e^{2\pi i \frac{k.n}{d}} \ , \quad n \in \Z_d \times \Z_d .
$$
It is easy to see by direct computation that the DCT intertwines translations with a phase factor, i.e.
$$
\wh{T(\lambda)f}(k) = e^{-2\pi i \frac{\lambda.k}{d}} \wh{f}(k) ,
$$
while it commutes with rotations, i.e.
\begin{equation}\label{eq:DFTrot}
\wh{Rf}(k_1,k_2) = \wh{f}(-k_2,k_1) = R \wh{f}(k_1,k_2).
\end{equation}
% $$
% \wh{Rf}(k_1,k_2) =
% \sum_{n_1 = - \frac{d-1}{2}}^{\frac{d-1}{2}} f(-n_2,n_1) e^{-2\pi i \frac{n_1k_1 + n_2k_2}{d}} , \quad k_1, k_2 \in \left\{-\frac{d-1}{2},\dots,\frac{d-1}{2}\right\}.
% \wh{f}(-k_2,k_1) = R \wh{f}(k_1,k_2)
% $$

\subsection{A group adapted isometry}\label{sec:isometry}

For a subgroup $\Lambda$ of $\Z_d \times \Z_d$, its annihilator is defined by
$$
L = \{\ell \in \Z_d \times \Z_d : e^{2\pi i \frac{\lambda.\ell}{d}} = 1 \ \ \forall \ \lambda \in \Lambda \} \subset \Z_d \times \Z_d.
$$
$L$ is a subgroup of $\Z_d \times \Z_d$ that plays a special role in Fourier analysis. The next lemma, whose proof is elementary, defines its structure and how it relates with rotations.

\begin{lem}\label{lem:annihilator}
Let $p, q$ be odd, let $d = pq$, and let $\Lambda$ be as in Lemma \ref{lem:subgroups}.
\begin{itemize}
\item[i)] The annihilator of $\Lambda$ is $L = \{\ell = (m_1,m_2)p : m_1, m_2 \in \{-\frac{q-1}{2},\dots,\frac{q-1}{2}\}\} \approx \Z_q \times \Z_q$.
\item[ii)] The set $\Omega = \{\omega = (\omega_1,\omega_2) : \omega_1,\omega_2 \in \{-\frac{p-1}{2},\dots,\frac{p-1}{2}\}\}\subset \Z_d \times \Z_d$ is a fundamental set for $(\Z_d \times \Z_d)/L$, in the sense that
\begin{itemize}
\item[.] $\Omega \cap (\Omega + \ell) = \emptyset$ for all $\ell \in L \setminus \{0\}$;
\item[.] $\Z_d \times \Z_d = \displaystyle\bigcup_{\ell \in L} (\Omega + \ell) = \Big\{k = \omega + \ell : \omega \in \Omega, \ell \in L\Big\}$.
\end{itemize}
\item[iii)] $L$ is invariant for $r$, i.e. $r(L) = L$, and for $\ell, \ell' \in L$ we have $r(\ell + \ell') = r(\ell) + r(\ell')$.
\item[iv)] Let $p > 1$. The set $\Omega_0 = \{\omega = (\omega_1,\omega_2) : \omega_1 \in \{1,\dots,\frac{p-1}{2}\}, \omega_2 \in \{0,\dots,\frac{p-1}{2}\}\}\subset \Omega$ satisfies
\begin{itemize}
\item[.] $\Omega_0 \cap r^g(\Omega_0) = \emptyset$ for all $g \in \{1,2,3\}$;
\item[.] $\Omega \setminus \{0\} = \displaystyle\bigcup_{g=0}^3 r^g(\Omega_0) = \Big\{\omega = r^g(\omega_0) : \omega_0 \in \Omega_0, g \in \{0,1,2,3\}\Big\}$.
\end{itemize}
\end{itemize}
\end{lem}

The following definition introduces a transform, denoted by $\T$, that is well adapted to perform Fourier analysis on $\Z_d \times \Z_d$ in the presence of the action of the group $\G$. This transform is a variation of the one introduced in Section 4.2 of [\citenum{BCHM2019}].
\begin{defi}
Let $f \in \ell_2(\Z_d \times \Z_d)$, with $d = pq$ odd, and let $L$ and $\Omega_0$ be as in Lemma \ref{lem:annihilator}. Let $N$ be the linear map on $\ell_2(\Z_d \times \Z_d)$ defined by
$$
N(f)(n) = \frac{1}{d^2} \left(\frac{1}{2}\sum_{\ell \in L} e^{2\pi i \frac{\ell . n}{d}} \wh{f}(\ell) + \sum_{k \in \Z_d \times \Z_d \smallsetminus L} e^{2\pi i \frac{k . n}{d}} \wh{f}(k) \right) \ , \quad n \in \Z_d \times \Z_d
$$
or, equivalently,
$
\wh{N(f)}(k) = \left\{
\begin{array}{rl}
\frac{1}{2} \wh{f}(k) \ ,& k \in L\vspace{1ex}\\
\wh{f}(k) \ ,& k \notin L .
\end{array}
\right.
$.
For any $\omega \in \Omega_0 \cup \{0\}$ we define
\begin{equation}\label{eq:TG}
\T [f](\omega) = \Big\{\frac{1}{d}\wh{R^g N(f)}(\omega + \ell) : g \in \{0,1,2,3\}, \ell \in L\Big\}
\end{equation}
which belongs to $\ell_2(\{0,1,2,3\} \times L)$. For each $g \in \{0,1,2,3\}$ we denote the corresponding element of $\ell_2(L)$ by
$$
\T [f](\omega)^g = \Big\{\T [f](\omega)^g_\ell : \ell \in L\Big\}
$$
where $\T [f](\omega)^g_\ell$ is the $(g,\ell)$ component of $\T [f](\omega)$:
$$
\T [f](\omega)^g_\ell = \frac{1}{d}\wh{R^g N(f)}(\omega + \ell) = \frac1d \left\{
\begin{array}{rl}
\displaystyle\frac12 \ \wh{R^g f}(\ell) \ ,& \omega = 0\vspace{1ex}\\
\wh{R^g f}(\omega + \ell) \ ,& \omega \neq 0
\end{array}
\right..
$$
\end{defi}

We provide now a proof of the following result, which also clarifies the role of the map $N$.
\begin{theo}\label{theo:isometry}
The map (\ref{eq:TG}) is an invertible isometry $\T : \ell_2(\Z_d \times \Z_d) \to \ell_2((\Omega_0 \cup \{0\}) \times \{0,1,2,3\} \times L)$. For $F \in \ell_2((\Omega_0 \cup \{0\}) \times \{0,1,2,3\} \times L)$, its inverse is
\begin{equation}\label{eq:inverseTG}
\T^{-1}[F](n)
\frac{1}{d}\sum_{g = 0}^3\left(\frac12 \sum_{\ell \in L} e^{2\pi i \frac{r^g(\ell) .n}{d}} F(0)^g_\ell + \sum_{\ell \in L} e^{2\pi i \frac{r^g(\ell) .n}{d}}\sum_{\omega \in \Omega_0} e^{2\pi i \frac{r^g(\omega) .n}{d}} F(\omega)^g_\ell\right).
\end{equation}
\end{theo}

\vspace{-2ex}
\noindent
\emph{Proof.}
Let us first prove that $\T$ is an isometry. Using (\ref{eq:DFTrot}) and Lemma \ref{lem:annihilator}
\begin{align*}
\sum_{\omega \in \Omega_0 \cup \{0\}} & \|\T [f](\omega)\|_{\ell_2(\{0,1,2,3\}\times L)}^2 = \frac{1}{d^2}\sum_{\omega \in \Omega_0 \cup \{0\}} \sum_{\ell \in L} \sum_{g \in \G} |\wh{R^g N(f)}(\omega + \ell)|^2\\
& = \frac{1}{d^2}\sum_{g \in \G} \sum_{\ell \in L} |\wh{R^g N(f)}(\ell)|^2
+ \frac{1}{d^2}\sum_{g \in \G} \sum_{\omega \in \Omega_0 } \sum_{\ell \in L} |\wh{R^g N(f)}(\omega + \ell)|^2\\
& = \frac{1}{d^2}\sum_{g \in \G} \sum_{\ell \in L} |\wh{N(f)}(\ell)|^2
+ \frac{1}{d^2}\sum_{g \in \G} \sum_{\omega \in \Omega_0 } \sum_{\ell \in L} |\wh{N(f)}(r^g\omega + \ell)|^2\\
& = \frac{1}{d^2}\left(4\sum_{\ell \in L} |\wh{N(f)}(\ell)|^2
+ \sum_{\omega \in \Omega \setminus \{0\} } \sum_{\ell \in L} |\wh{N(f)}(\omega + \ell)|^2\right) = \frac{1}{d^2}\sum_{\omega \in \Omega} \sum_{\ell \in L} |\wh{f}(\omega + \ell)|^2\\
& = \frac{1}{d^2}\|\wh{f}\|_d^2 = \|f\|_d^2 .
\end{align*}

To prove the inversion formula, again using Lemma \ref{lem:annihilator}
\begin{align*}
f(n) & = \frac{1}{d^2}\sum_{k \in \Z_d \times \Z_d} e^{2\pi i \frac{k.n}{d}} \wh{f}(k) = \frac{1}{d^2}\sum_{\ell \in L} e^{2\pi i \frac{\ell .n}{d}}\sum_{\omega \in \Omega} e^{2\pi i \frac{\omega .n}{d}} \wh{f}(\omega + \ell)\\
& = \frac{1}{d^2}\sum_{\ell \in L} e^{2\pi i \frac{\ell .n}{d}} \wh{f}(\ell) + \frac{1}{d^2}\sum_{\ell \in L} e^{2\pi i \frac{\ell .n}{d}}\sum_{g = 0}^3\sum_{\omega \in \Omega_0} e^{2\pi i \frac{r^g(\omega) .n}{d}} \wh{f}(r^g(\omega) + \ell)\\
& = \frac{1}{d^2}\sum_{g = 0}^3\left(\frac14 \sum_{\ell \in L} e^{2\pi i \frac{r^g(\ell) .n}{d}} \wh{f}(r^g(\ell)) + \sum_{\ell \in L} e^{2\pi i \frac{r^g(\ell) .n}{d}}\sum_{\omega \in \Omega_0} e^{2\pi i \frac{r^g(\omega) .n}{d}} \wh{f}(r^g(\omega + \ell))\right)\\
& = \frac{1}{d^2}\sum_{g = 0}^3\left(\frac12 \sum_{\ell \in L} e^{2\pi i \frac{r^g(\ell) .n}{d}} \wh{N(f)}(r^g(\ell)) + \sum_{\ell \in L} e^{2\pi i \frac{r^g(\ell) .n}{d}}\sum_{\omega \in \Omega_0} e^{2\pi i \frac{r^g(\omega) .n}{d}} \wh{N(f)}(r^g(\omega + \ell))\right)\\
& = \frac{1}{d}\sum_{g = 0}^3\left(\frac12 \sum_{\ell \in L} e^{2\pi i \frac{r^g(\ell) .n}{d}} \T[f](0)^g_\ell + \sum_{\ell \in L} e^{2\pi i \frac{r^g(\ell) .n}{d}}\sum_{\omega \in \Omega_0} e^{2\pi i \frac{r^g(\omega) .n}{d}} \T[f](\omega)^g_\ell\right)
\end{align*}
where the last identity is due to (\ref{eq:DFTrot}). \hfill $\Box$

The next lemma shows how the isometry $\T$ intertwines the action of $\G$.

\begin{lem}\label{lem:intertwining}
Let $\Lambda$ be as in Lemma \ref{lem:subgroups}, and let $L$ and $\Omega_0$ be as in Lemma \ref{lem:annihilator}.
For all $f \in \ell_2(\Z_d \times \Z_d)$, all $(\omega,g,\ell) \in (\Omega_0 \cup \{0\}) \times \{0,1,2,3\} \times L$, all $\lambda \in \Lambda$ and all $g' \in \{0,1,2,3\}$ it holds
$$
\T [T(\lambda)f](\omega)^g = e^{-2\pi i \frac{\lambda.r^g(\omega)}{d}} \T [f](\omega)^g
$$
and
$$
\T [R^{g'}\! f](\omega)^g = \T [f](\omega)^{g+g'}.
$$
\end{lem}

\vspace{-2ex}
\noindent
\emph{Proof.}
Observe first that, since $N$ is only a normalization of the Fourier coefficients on the subgroup $L$, it commutes with translations: $N(T(\lambda)f) = T(\lambda)N(f)$. Thus, for all $\ell \in L$
\begin{align*}
\T [T(\lambda)f](\omega)_\ell^g & = \frac{1}{d}\Big(R^g N(T(\lambda)f)\Big)^\wedge(\omega + \ell) = \frac{1}{d}\Big( N(T(\lambda)f)\Big)^\wedge(r^g(\omega + \ell))\\
& = \frac{1}{d}\Big( T(\lambda)N(f)\Big)^\wedge(r^g(\omega + \ell)) = \frac{1}{d}e^{-2\pi i \frac{\lambda.r^g(\omega)}{d}}\wh{ N(f)}(r^g(\omega + \ell))\\
& = e^{-2\pi i \frac{\lambda.r^g(\omega)}{d}} \T[f](\omega)_\ell^g ,
\end{align*}
where the second to last identity makes use of the fact that $L$ is the annihilator of $\Lambda$ and of the invariance of $L$ under rotations. This last fact also implies that $N$ commutes with rotations: $RN(f) = N(Rf)$. Thus
$$
\hspace{124pt}\T [R^{g'} f](\omega)_\ell^g = \frac{1}{d}\Big(R^g N(R^{g'}f)\Big)^\wedge(\omega + \ell) = \T [f](\omega)_\ell^{g+g'}. \hspace{114pt}\Box
$$

\subsection{Group invariant spaces}

Using the previous results, we can deduce how $\G$-invariant spaces are transformed under the action of the map $\T$. The next theorem shows that, for any fixed $(\omega,g) \in (\Omega_0 \cup \{0\}) \times \{0,1,2,3\}$, the $\ell_2(L)$ element obtained by the $\T$ transform of a linear combination of translates and rotates of a family $\Psi = \{\psi_j\}_{j= 1}^\kappa \subset \Z_d \times \Z_d$ is a linear combination of the $\T$ transform of the rotates of $\Psi$.
\begin{theo}\label{theo:GSIS}
Let $\Lambda$ be as in Lemma \ref{lem:subgroups}, and let $L$ and $\Omega_0$ be as in Lemma \ref{lem:annihilator}. Let $\Psi = \{\psi_j\}_{j= 1}^\kappa \subset \Z_d \times \Z_d$, let $S(\Psi)$ be as in Definition \ref{def:GSIS}, and let $\T$ be the map defined in (\ref{eq:TG}). For $\omega \in \Omega_0 \cup \{0\}$ and $g \in \{0,1,2,3\}$, let us denote by
$$
\T[S(\Psi)](\omega)^g = \Big\{v = \T[f](\omega)^g : f \in S(\Psi) \Big\}
$$
and let us denote by
\begin{equation}\label{eq:rangefunction}
\mathscr{S}_\Psi(\omega) = \vsp\Big\{\T[\psi_j](\omega)^{g'} : j \in \{1,\dots,\kappa\}, g' \in \{0,1,2,3\}\Big\} .
\end{equation}
Then these two subspaces of $\ell_2(L)$ coincide:
\begin{equation}\label{eq:range}
\T[S(\Psi)](\omega)^g = \mathscr{S}_\Psi(\omega).
\end{equation}
\end{theo}

\vspace{-2ex}
\noindent
\emph{Proof.}
Observe first that (\ref{eq:range}) implies in particular that for any $f \in S(\Psi)$, any $g, g' \in \{0,1,2,3\}$, and any $\omega \in \Omega_0 \cup \{0\}$, the elements $\T[f](\omega)^g$ and $\T[f](\omega)^{g'}$ belong to the same subspace of $\ell_2(L)$. This, however, can already be deduced by Lemma \ref{lem:intertwining}, because $\T[f](\omega)^g = \T[R^{g-g'}f](\omega)^{g'}$, and $S(\Psi)$ is invariant under rotations.

Let us now prove that $\T[S(\Psi)](\omega)^g \subset \mathscr{S}_\Psi(\omega)$ for every $\omega \in \Omega_0 \cup \{0\}$ and every $g \in \{0,1,2,3\}$, noting that, by the previous argument, it suffices to prove this only for $g = 0$.

Let $f \displaystyle = \sum_{\lambda \in \Lambda} \sum_{g' = 0}^3 \sum_{j= 1}^\kappa c_j(\lambda,g') T(\lambda)R^{g'}\psi_j \in S(\Psi)$. Then, using Lemma \ref{lem:intertwining}
% \begin{align*}
% \T[f](\omega)^g & = \sum_{g' = 0}^3 \sum_{j= 1}^\kappa \sum_{\lambda \in \Lambda} c_j(\lambda,g') \T[T(\lambda)R^{g'}\psi_j](\omega)^g\\
% & = \sum_{g' = 0}^3 \sum_{j= 1}^\kappa \left(\sum_{\lambda \in \Lambda} c_j(\lambda,g') e^{-2\pi i \frac{\lambda.r^g(\omega)}{d}}\right)\T[\psi_j](\omega)^{g + g'}\\
% & = \sum_{g'' = 0}^3 \sum_{j= 1}^\kappa \left(\sum_{\lambda \in \Lambda} c_j(\lambda,g''-g) e^{-2\pi i \frac{\lambda.r^g(\omega)}{d}}\right)\T[\psi_j](\omega)^{g''}.
% \end{align*}
% Let us denote by
% $$
% C_j(\omega,g') = \sum_{\lambda \in \Lambda} c_j(\lambda,g') e^{-2\pi i \frac{\lambda.\omega}{d}}
% $$
% and let $A^g : C_j(\omega,g') \mapsto A^gC_j(\omega,g') = C_j(r^g(\omega),g'-g)$. Then
% $$
% \T[f](\omega)^g = \sum_{g'' = 0}^3 \sum_{j= 1}^\kappa A^gC_j(\omega,g'') \T[\psi_j](\omega)^{g''} \in \mathscr{S}_\Psi(\omega).
% $$
\begin{align}\label{eq:TGfinSPsi}
\T[f](\omega)^0 & = \sum_{g' = 0}^3 \sum_{j= 1}^\kappa \sum_{\lambda \in \Lambda} c_j(\lambda,g') \T[T(\lambda)R^{g'}\psi_j](\omega)^0 = \sum_{g' = 0}^3 \sum_{j= 1}^\kappa \bigg(\sum_{\lambda \in \Lambda} c_j(\lambda,g') e^{-2\pi i \frac{\lambda.\omega}{d}}\bigg)\T[\psi_j](\omega)^{g'}\nonumber\\
& = \sum_{g' = 0}^3 \sum_{j= 1}^\kappa C_j(\omega,g') \T[\psi_j](\omega)^{g'} \in \mathscr{S}_\Psi(\omega)
\end{align}
where we have set
\begin{equation}\label{eq:GammaFourier}
C_j(\omega,g') = \sum_{\lambda \in \Lambda} c_j(\lambda,g') e^{-2\pi i \frac{\lambda.\omega}{d}}.
\end{equation}
This proves that $\T[S(\Psi)](\omega)^g \subset \mathscr{S}_\Psi(\omega)$ for every $\omega \in \Omega_0 \cup \{0\}$ and every $g \in \{0,1,2,3\}$.

To prove the opposite inclusion, fix $\omega_0 \in \Omega_0 \cup \{0\}$ and let $v \displaystyle = \sum_{g' = 0}^3 \sum_{j= 1}^\kappa a_j(g') \T[\psi_j](\omega_0)^{g'} \in \mathscr{S}_\Psi(\omega_0)$. We want to prove that for each $g \in \{0,1,2,3\}$ there exists $f_v \in S(\Psi)$ such that $\T[f_v](\omega_0)^g = v$. Again, since $S(\Psi)$ is rotation invariant, by Lemma \ref{lem:intertwining} it suffices to consider only $g = 0$. Now, observe that (\ref{eq:GammaFourier}) is a DFT for $\Z_p \times \Z_p$. Indeed since $\lambda = (\lambda_1,\lambda_2) = (t_1,t_2)q = tq$ for $t \in \Z_p \times \Z_p$, equation (\ref{eq:GammaFourier}) can be written as $C_j(\omega,g') = \displaystyle\sum_{t \in \Z_p \times \Z_p} c_j(t q,g') e^{-2\pi i \frac{t.\omega}{p}}$. Hence, it is inverted by
$$
c_j(\lambda,g') = \frac{1}{p^2}\displaystyle\sum_{\omega \in \Omega} C_j(\omega,g') e^{2\pi i \frac{\lambda.\omega}{d}}.
$$
Thus, for
$
C_j(\omega,g') = \left\{
\begin{array}{cc}
a_j(g') & \omega = \omega_0\\
0 & \omega \neq \omega_0
\end{array}
\right.
$
we get $c_j(\lambda,g') = \displaystyle \frac{1}{p^2}a_j(g')e^{2\pi i \frac{\lambda.\omega_0}{d}}$. Now, by (\ref{eq:TGfinSPsi}), the element of $S(\Psi)$ given by
$$
f_v = \frac{1}{p^2}\sum_{\lambda \in \Lambda} \sum_{g = 0}^3 \sum_{j= 1}^\kappa a_j(g')e^{2\pi i \frac{\lambda.\omega_0}{d}} T(\lambda)R^{g'}\psi_j
$$
satisfies $\T[f_v](\omega_0)^0 = v$. This concludes the proof. \hfill $\Box$

\begin{rem}
The function $\omega \mapsto \mathscr{S}_\Psi(\omega)$ is a nonabelian variant of an object known in the literature on shift-invariant spaces as \emph{range function} (see e.g. [\citenum{Bownik2000, CP2010}] and references therein), and, up to a minor change, it corresponds to the map introduced in Definition 4.5 of [\citenum{BCHM2019}].
We note here also that, for the same $f \in S(\Psi)$ used in the previous proof, we can easily compute the $g$ components of $\T[f](\omega)$. Using Lemma \ref{lem:intertwining} and (\ref{eq:TGfinSPsi})
$$
\T[f](\omega)^g = \sum_{g' = 0}^3 \sum_{j= 1}^\kappa C_j(r^g(\omega),g'-g) \T[\psi_j](\omega)^{g'} \ \in \mathscr{S}_\Psi(\omega).
$$
Indeed, $\T[f](\omega)^g = \T[R^gf](\omega)^0$, so the previous formula can be deduced using (\ref{eq:commutator}).
% A proof is:
% \begin{align*}
% \T[f](\omega)^g & = \T[R^gf](\omega)^0 = \sum_{g' = 0}^3 \sum_{j= 1}^\kappa \sum_{\lambda \in \Lambda} c_j(\lambda,g') \T[R^gT(\lambda)R^{g'}\psi_j](\omega)^0\\
% & = \sum_{g' = 0}^3 \sum_{j= 1}^\kappa \sum_{\lambda \in \Lambda} c_j(\lambda,g') \T[T(r^{-g}\lambda)R^{g+g'}\psi_j](\omega)^0 = \sum_{g'' = 0}^3 \sum_{j= 1}^\kappa \sum_{\lambda \in \Lambda} c_j(r^g\lambda,g''-g) \T[T(\lambda)R^{g''}\psi_j](\omega)^0 .
% \end{align*}
% % In particular, the new coefficients are obtained via the action of the rotation group on $\Omega \times \{0,1,2,3\}$ defined by $A^g : (\omega,g') \mapsto A^g(\omega,g') = (r^g(\omega),g'-g)$.
\end{rem}

\newpage
The next result is the analogous of a lemma by Helson that is crucial in many aspects of the theory of shift-invariant spaces. In the present form, it is a minor variation of Lemma 4.6 of [\citenum{BCHM2019}], whose proof can be obtained following similar arguments to those in Proposition 3.9 of [\citenum{CP2010}]. However, in this case can obtain it as a corollary of Theorem \ref{theo:GSIS}.

\begin{cor}\label{cor:projections}
Let $\Psi = \{\psi_j\}_{j= 1}^\kappa \subset \Z_d \times \Z_d$, and let $\T$ be the map defined in (\ref{eq:TG}). For $S(\Psi)$ as in Definition \ref{def:GSIS}, let $\Proj_{S(\Psi)}$ denote the orthogonal projection of $\ell_2(\Z_d \times \Z_d)$ onto $S(\Psi)$. For $\mathscr{S}_\Psi$ as in (\ref{eq:rangefunction}) and $\omega \in \Omega_0 \cup \{0\}$, let $\pj_{\mathscr{S}_\Psi(\omega)}$ denote the orthogonal projection of $\ell_2(L)$ onto $\mathscr{S}_\Psi(\omega)$. Then, for all $g \in \{0,1,2,3\}$,
$$
\T[\Proj_{S(\Psi)}f](\omega)^g = \pj_{\mathscr{S}_\Psi(\omega)}\T[f](\omega)^g \quad \forall \ f \in \ell_2(\Z_d \times \Z_d).
$$
\end{cor}

\vspace{-2ex}
\noindent
\emph{Proof.}
Let $Q_1$ be the orthogonal projection of $\ell_2((\Omega_0 \cup \{0\}) \times \{0,1,2,3\} \times L)$ onto $\T[S(\Psi)]$,
% defined by
% $$
% Q_1 F = \T \big[\Proj_{S(\Psi)} \T^{-1}[F]\big] \quad \forall \ F \in \ell_2((\Omega_0 \cup \{0\}) \times \{0,1,2,3\} \times L)
% $$
and let $Q_2$ be the linear operator on $\ell_2((\Omega_0 \cup \{0\}) \times \{0,1,2,3\} \times L)$ defined by
$$
Q_2 F (\omega)^g = \pj_{\mathscr{S}_\Psi(\omega)} F(\omega)^g \quad \forall \ F \in \ell_2((\Omega_0 \cup \{0\}) \times \{0,1,2,3\} \times L) , \ \forall (\omega,g) \in (\Omega_0 \cup \{0\}) \times \{0,1,2,3\}.
$$
Note that $Q_2$ is also an orthogonal projection, because $Q_2^2 = Q_2^* = Q_2$. Moreover, by definition
$$
Q_1\T[f](\omega)^g = \T[\Proj_{S(\Psi)}f](\omega)^g \quad \forall \ f \in \ell_2(\Z_d \times \Z_d) , \ \forall (\omega,g) \in (\Omega_0 \cup \{0\}) \times \{0,1,2,3\}.
$$
The statement is then proved by showing that $Q_2$ and $Q_1$ have the same range. By Theorem \ref{theo:GSIS}, we have that $\textnormal{ran}(Q_1) \subset \textnormal{ran}(Q_2)$, because $\T[\Proj_{S(\Psi)}f](\omega)^g \in \mathscr{S}_\Psi(\omega)$ for all $(\omega,g) \in (\Omega_0 \cup \{0\}) \times \{0,1,2,3\}$. For the opposite inclusion, observe that, again by Theorem \ref{theo:GSIS}, we have
\begin{align*}
\textnormal{ran}(Q_2) & = \Big\{F \in \ell_2((\Omega_0 \cup \{0\}) \times \{0,1,2,3\} \times L) : F(\omega)^g \in \mathscr{S}_\Psi(\omega) \ \ \forall \ (\omega,g) \in (\Omega_0 \cup \{0\}) \times \{0,1,2,3\}\Big\} \\
& = \Big\{F \in \ell_2((\Omega_0 \cup \{0\}) \times \{0,1,2,3\} \times L) : F(\omega)^g \in \T[S(\Psi)](\omega)^g \ \ \forall \ (\omega,g) \in (\Omega_0 \cup \{0\}) \times \{0,1,2,3\}\Big\} .
\end{align*}
This proves that $\textnormal{ran}(Q_2) \subset \textnormal{ran}(Q_1)$. \hfill $\Box$

\section{Solution to the approximation problem}\label{sec:solution}

In this section we provide a solution to the invariant approximation problem stated in Section \ref{sec:problem} for a dataset $\data = \{f_1,\dots,f_m\} \subset \ell_2(\Z_d \times \Z_d)$. The core idea is that the isometry introduced in Section \ref{sec:isometry} allows us to reduce the problem, which is stated in a space of dimension $d^2 = p^2q^2$ and contains $\G$-invariance constraints, to a sequence of problems in the smaller space $\ell_2(L)$, of dimension $q^2$, without invariance constraints, for datasets obtained by data augmentation with rotations. The datasets in the reduced problems are obtained via the map (\ref{eq:TG}), considering each $\omega \in \Omega_0 \cup \{0\}$ separately, so the proposed solution amounts to solve $\sharp \Omega_0 + 1 = \frac{p^2-1}{4} + 1$ quadratic optimization problems, which can be addressed with the ordinary SVD. The data augmentation that appears in this solution is due to the structure of (\ref{eq:TG}): as a consequence of Lemma \ref{lem:intertwining}, for $f \in \ell_2(\Z_d \times \Z_d)$, we have that $\T[f](\omega) = \{\T[R^gf](\omega)^0 : g \in \{0,1,2,3\}\}$, which means that in the reduced problems the original data are considered together with their rotates. We will first state the main result in the form of a theorem, then outline the structure of the algorithm that provides a solution in the form of the generators of an optimal invariant approximation subspace.

\begin{theo}
Let $p,q$ be odd, let $d = pq$, let $\data = \{f_1, \dots, f_m\} \subset \ell_2(\Z_d \times \Z_d)$ and let $\kappa \leq \min\{m,\frac{q^2}{4}\}$. Let $L$ and $\Omega_0$ be as in Lemma \ref{lem:annihilator}, and let $\T$ be the isometry (\ref{eq:TG}). For $\omega \in \Omega_0 \cup \{0\}$, define
\begin{equation}\label{eq:rdata}
\rdata_\omega = \{x_\omega^{i,g} = \T[f_i](\omega)^g : i \in \{1,\dots,m\}, g \in \{0,1,2,3\}\} \subset \ell_2(L).
\end{equation}
% The problem of minimizing (\ref{eq:error}) with a $\Gamma$-invariant subspace of $\ell_2(\Z_d \times \Z_d)$ defined by $\kappa$ Parseval frame generators can be solved by minimizing
The minimization problem (\ref{eq:error}) can be solved by minimizing, at each $\omega \in \Omega_0 \cup \{0\}$,
\begin{equation}\label{eq:reduced}
\D[\rdata_\omega,\W_\omega] = \sum_{i=1}^m \sum_{g = 0}^3 \|x_\omega^{i,g} - \pj_{\W_\omega}x_\omega^{i,g}\|_{\ell_2(L)}^2
\end{equation}
over the linear subspaces $\W_\omega \subset \ell_2(L)$ of dimension at most $\kappa$, where $\pj_{\W_\omega}$ denotes the orthogonal projection of $\ell_2(L)$ onto $\W_\omega$.
\end{theo}

\newpage
\noindent
\emph{Proof.}
The main computation for this proof is the following, which makes use of Theorem \ref{theo:isometry} and Corollary \ref{cor:projections}:
\begin{align}\label{eq:reduction}
\sum_{i=1}^m \|f_i - \Proj_{S(\Psi)}f_i\|^2 & = \sum_{\omega \in \Omega_0 \cup \{0\}} \sum_{i=1}^m \sum_{g = 0}^3 \|\T[f_i](\omega)^g - \T[\Proj_{S(\Psi)}f_i](\omega)^g\|_{\ell_2(L)}^2\nonumber\\
& = \sum_{\omega \in \Omega_0 \cup \{0\}} \sum_{i=1}^m \sum_{g = 0}^3 \|\T[f_i](\omega)^g - \pj_{\mathscr{S}_\Psi(\omega)}\T[f_i](\omega)^g\|_{\ell_2(L)}^2 = \sum_{\omega \in \Omega_0 \cup \{0\}} \D[\rdata_\omega,\mathscr{S}_\Psi(\omega)].
\end{align}
Now, suppose we consider separately, for each $\omega \in \Omega_0 \cup \{0\}$, the reduced problem (\ref{eq:reduced}). Let $X_\omega$ be the matrix containing the data $\rdata_\omega \subset \ell_2(L)$ defined in (\ref{eq:rdata}) organized by columns
\begin{equation}\label{eq:matrix}
X_\omega = 
\underbrace{
\left(
\begin{array}{ccccccccc}
\vdots & \vdots & \vdots & \vdots & & \vdots & \vdots & \vdots & \vdots \\
&&&&&&&&\\
x_\omega^{1,0} & x_\omega^{1,1} & x_\omega^{1,2} & x_\omega^{1,3} & \dots & x_\omega^{m,0} & x_\omega^{m,1} & x_\omega^{m,2} & x_\omega^{m,3}\\
&&&&&&&&\\
\vdots & \vdots & \vdots & \vdots & & \vdots & \vdots & \vdots & \vdots
\end{array}
\right)
}_{4m}
\hspace{-40pt}
\left.
\phantom{
\left(
\begin{array}{c}
\vdots \\
\\
x_\omega^{1,0}\\
\\
\vdots
\end{array}
\right)
}
\right\} q^2
\end{equation}
and let
$$
X_\omega = U_\omega \Sigma_\omega V_\omega^*
$$
be its SVD. In the typical case of $4m > q^2$, we have that $U_\omega \in \C^{q^2 \times q^2}$ is a unitary matrix, and $\Sigma_\omega$ is a diagonal matrix $\in \R_+^{q^2 \times q^2}$ whose diagonal entries $\{\sigma_i\}_{i=1}^{q^2}$ are ordered decreasingly. For $\kappa \leq \min\{m,\frac{q^2}{4}\}$, let $\{U_\omega^s\}_{s=1}^{4\kappa}$ be the first $4\kappa$ columns of $U_\omega$ re-organized as elements of $\ell_2(L)$, and let
$$
\V_\omega = \vsp\{U_\omega^s\}_{s=1}^{4\kappa} \subset \ell_2(L) .
$$
Then $\V_\omega$ is a minimizer of (\ref{eq:reduced}), because for all subspaces $\W_\omega \subset \ell_2(L)$ of dimension at most $4\kappa$, we have
$$
\D[\rdata_\omega,\V_\omega] \leq \D[\rdata_\omega,\W_\omega].
$$
Let us now define the elements $\{\varphi_j(\omega)^g : j \in \{1,\dots,\kappa\}, g \in \{0,1,2,3\}\} \subset \ell_2(L)$ as follows:
\begin{equation}\label{eq:CHOICE}
\varphi_j(\omega)^g = U_\omega^{4(j-1)+g+1} \ , \quad j \in \{1,\dots,\kappa\}, g \in \{0,1,2,3\} .
\end{equation}
By performing this construction for each $\omega \in \Omega_0 \cup \{0\}$, we then get a family
$$
\{\varphi_j\}_{j = 1}^\kappa \subset \ell_2((\Omega_0 \cup \{0\}) \times \{0,1,2,3\} \times L).
$$
Let now $\Phi = \{\phi_j\}_{j=1}^\kappa \subset \ell_2(\Z_d \times \Z_d)$ be obtained by applying (\ref{eq:inverseTG}):
\begin{equation}\label{eq:SOLUTION}
\phi_j = \T^{-1}[\varphi_j] , \quad j \in \{1,\dots,\kappa\}.
\end{equation}
Then, by construction, $\mathscr{S}_\Phi(\omega) = \V_\omega$ for each $\omega \in \Omega_0 \cup \{0\}$, and, by (\ref{eq:reduction}), the family $\Phi$ is a minimizer for (\ref{eq:error}). The proof that the translates and rotates of $\Phi$ actually form a Parseval frame is left to the reader, whom we invite to refer to the argument provided in [\citenum{BCHM2019}].
\hfill $\Box$

\begin{rem}
We observe that the solution provided is not the unique one satisfying the desired constraints. Indeed, first of all it is well-known that, in case of degeneracies in the singular values, the approximation provided by the SVD may not be unique, and also in case of no degeneracies it is unique only up to a unitary transformation. However, also the choice of global generators that is performed in (\ref{eq:CHOICE}) is not the only possible choice. Other orderings of the columns of $U_\omega$, eventually also depending on $\omega$, could be chosen instead of (\ref{eq:CHOICE}), which would define a different family of Parseval frame generators for the same optimal invariant space.
\end{rem}

\newpage
We can now outline as a pseudocode the constructive scheme that produces the solution given in the previous proof. Recall that we consider a dataset of $d \times d$ images $\data = \{f_1,\dots,f_m\}$, with $d = pq$ for $p,q$ odd, and with $L$ and $\Omega_0$ as in Lemma \ref{lem:annihilator}.

\begin{center}
\hspace{24pt}
\begin{minipage}[h!]{.85\textwidth}
\begin{lstlisting}[numbers = left]
for $\omega \in \Omega_0 \cup \{0\}$
    for $i \in \{1,\dots,m\}, g \in \{0,1,2,3\}$ 
        compute $x^{i,g} = \T[f_i](\omega)^g$ using $(3.2)$ 
    end 
    organize $\{x^{i,g} : i \in \{1,\dots,m\}, g \in \{0,1,2,3\}\}$ in a matrix $X$ as in $(4.3)$ 
    compute the first $4\kappa$ columns of $U$ in the SVD of $X = U\Sigma V^*$
    re-organize them into elements $\{U^s\}_{s = 1}^{4\kappa}$ of $\ell_2(L)$
    for $j \in \{1,\dots,\kappa\}, g \in \{0,1,2,3\}$ 
        store $\varphi_j(\omega)^g = U^{4(j-1)+g+1}$ 
    end
end
for $j \in \{1,\dots,\kappa\}$ 
    compute $\phi_j = \T^{-1}[\varphi_j]$ using $(3.3)$
end
\end{lstlisting}
\centering
\vspace{.5ex}
Table 1: Pseudocode to compute the generators for the $\Gamma$-invariant approximation of $\data$\hspace{24pt}.
\end{minipage}
\end{center}

Most of this construction actually concerns the computations in lines 2-4 and in line 13. We want to spend a few words here regarding their implementation.

First of all, the fastest way to implement line 13 is certainly not to use directly (\ref{eq:inverseTG}), but rather to first recast all components of each $\varphi_j$ into the full Fourier transform of $\phi_j$, and then perform the inversion by FFT. This, however, will not have a significant impact on the overall performances whenever the number $\kappa$ of generators to be computed is small.

On the other hand, for lines 2-4, we have to consider two main issues. The first one is that no relevant performance improvement can be expected by replacing the full FFT with partial versions, so line 3 can be implemented as essentially a selection of FFT coefficients previously computed. The second issue is that, if we wish to consider a large lattice $\Lambda$ of translations, then also $\Omega$ will be large, because they have the same number of points. In this case, line 3 in the loop in $\omega$ requires to compute a number of FFT of images corresponding to a large multiple of the size of the dataset. This may be very time consuming, and one may rather wish to first compute the full FFT of the whole dataset, and then at each step of the loop in $\omega$ just perform the coefficients selection required by line 3. However, this approach requires to store the FFT of the whole dataset at once. If one aims to deal with a large dataset, this can be very demanding at the level of memory availability. Indeed, for a greyscale digital image, the values that a pixel can take are typically stored in 1 byte (8 bits). On the other hand, its FFT coefficients are complex floating point, so each one requires 16 bytes. Hence, for a given amount of available memory, the approach of computing a large number of FFT allows one to deal with a dataset that can be 16 times larger than the one that could be processed by first storing the FFT of the whole dataset. For large datasets, with $4m \gg q^2$, one may consider instead to first build $X^{\phantom{*}}_\omega X_\omega^*$ incrementally for each $\omega$ by splitting the dataset, store them, and then iterate again over $\omega$ lines 6-10.

\newpage

\section{Numerical results}\label{sec:numerics}

We present a test of our method on the ImageNet ILSVRC2017\footnote{\url{http://image-net.org/challenges/LSVRC/2017/}} dataset [\citenum{ILSVRC15}]. In this dataset, images have different sizes: we have cropped them to $d \times d$ pixels, and converted to 8 bits grayscale when necessary. For convenience, we have also removed the average of the dataset from each image before starting the processing.

The numerical results are presented for $d = 345$, with $p = 23$ and $q = 15$. Recall that, with the notation of Lemma \ref{lem:subgroups}, the number of points in the lattice $\Lambda$ is $p^2$, while $q$ indicates the spacing between the points of $\Lambda$. The size of the dataset we have used is $m = 2000$ images. The upper bound for the number $\kappa$ of generators that we want to consider is $\displaystyle\big[q^2/4\big] = 56$: we will show the results for values $\kappa = 8, 14$ and $19$, which correspond to subspaces of dimension at most $\frac17, \frac14$ and $\frac13$ of the dimension of $\ell_2(\Z_d \times \Z_d)$ (see the discussion in Section \ref{sec:problem}).

\begin{figure}[h!]
\begin{center}
\includegraphics[width=0.3\textwidth]{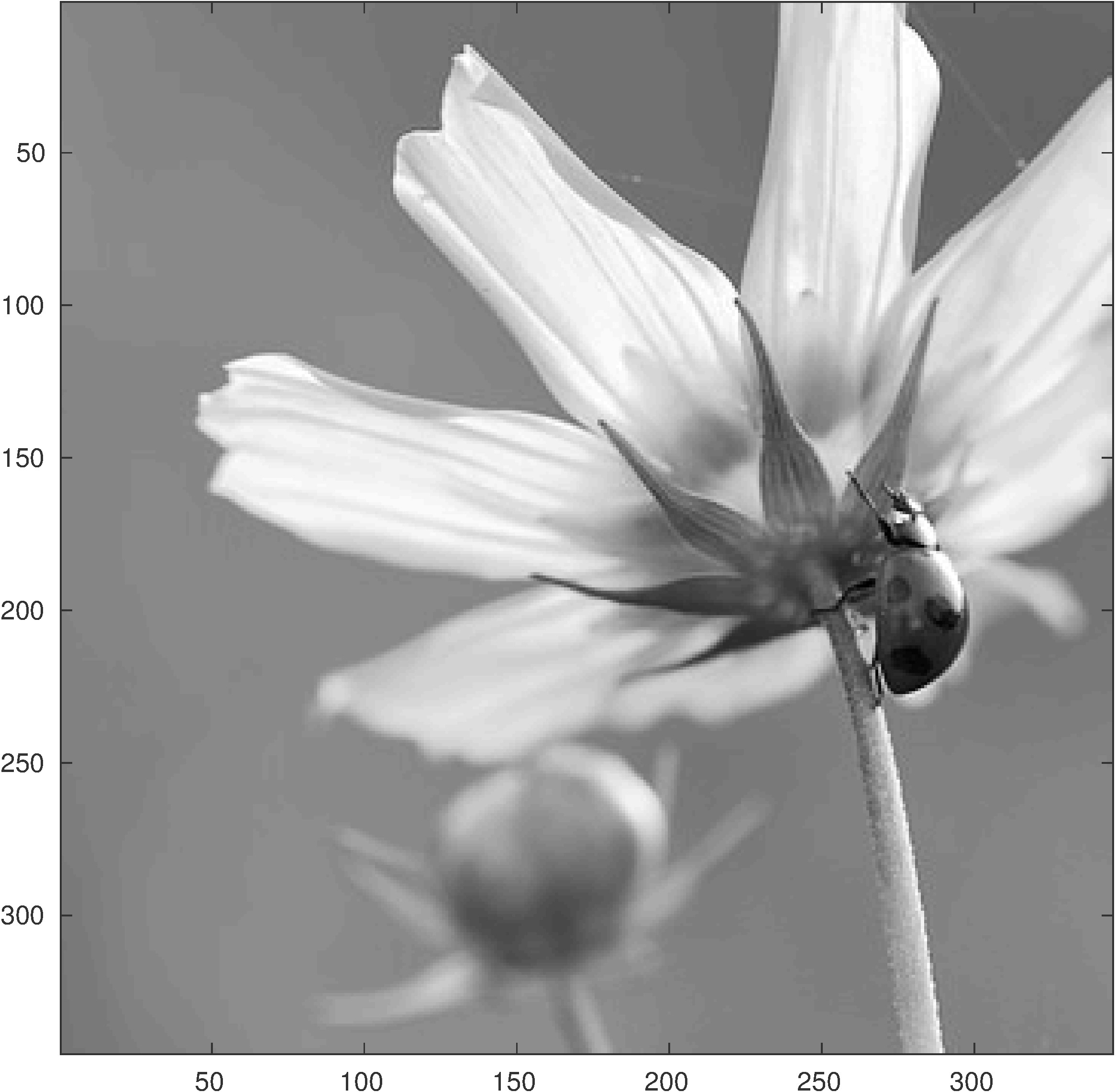} \qquad
\includegraphics[width=0.3\textwidth]{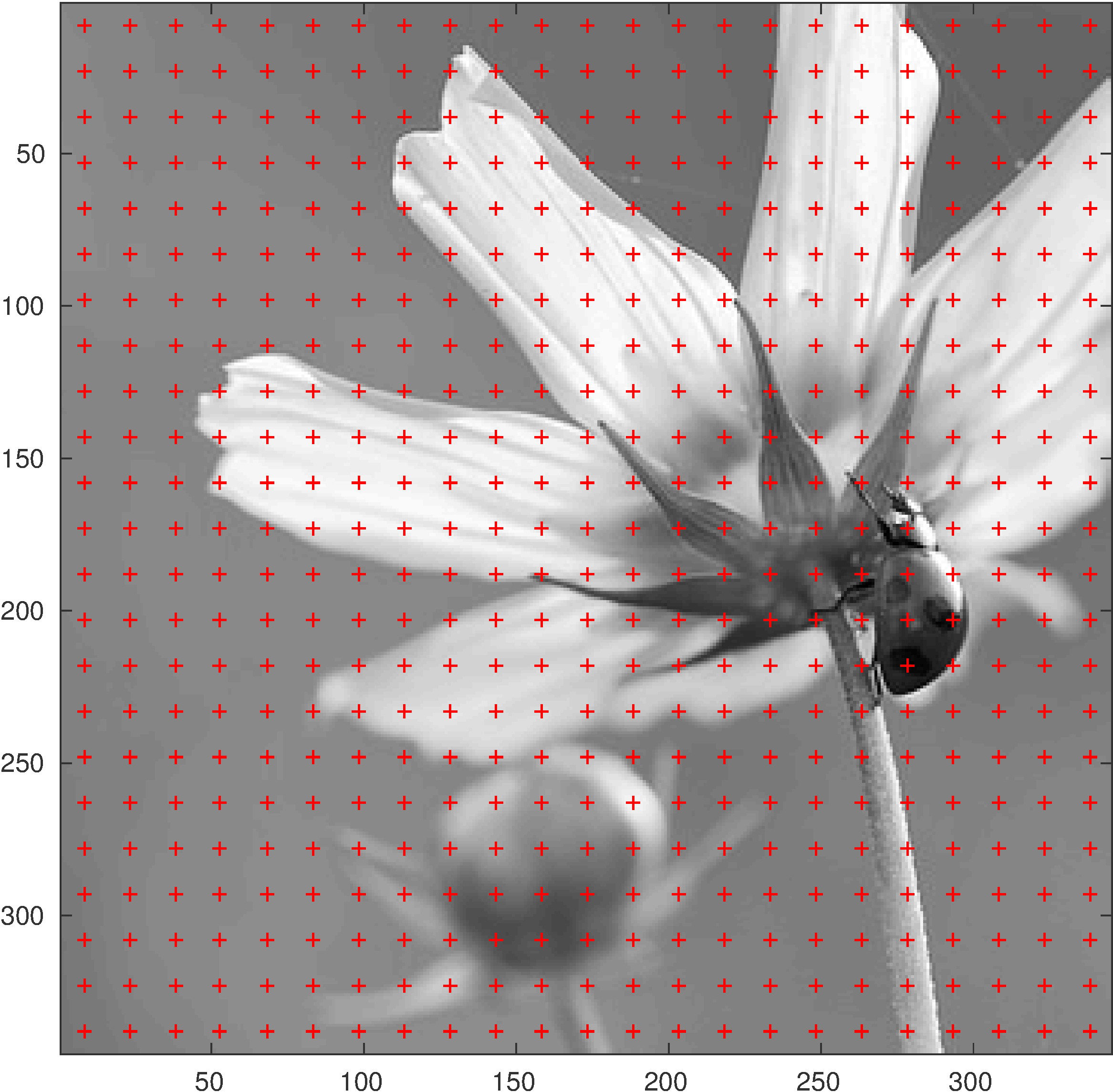} \qquad
\includegraphics[width=0.3\textwidth]{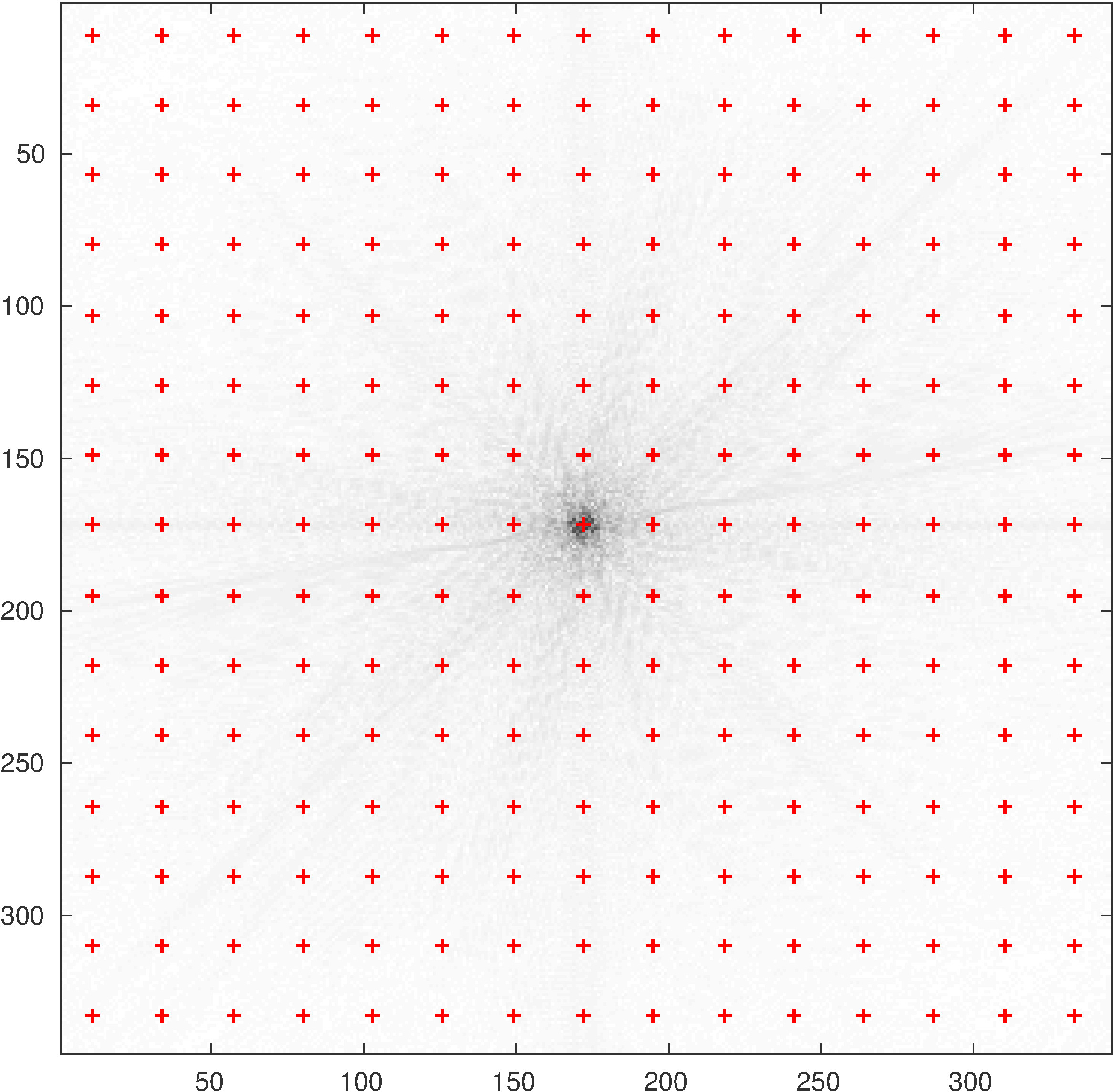}
\end{center}
\caption{\label{fig:lattice} 
Left: matrix visualization of an image in the dataset. Center: lattice $\Lambda$ of translates for $p = 23$ and $q = 15$. Right: modulus of the Fourier transform of the image and dual lattice $L$.}
\end{figure}

\begin{figure}[h!]
\begin{center}
\includegraphics[width=\textwidth]{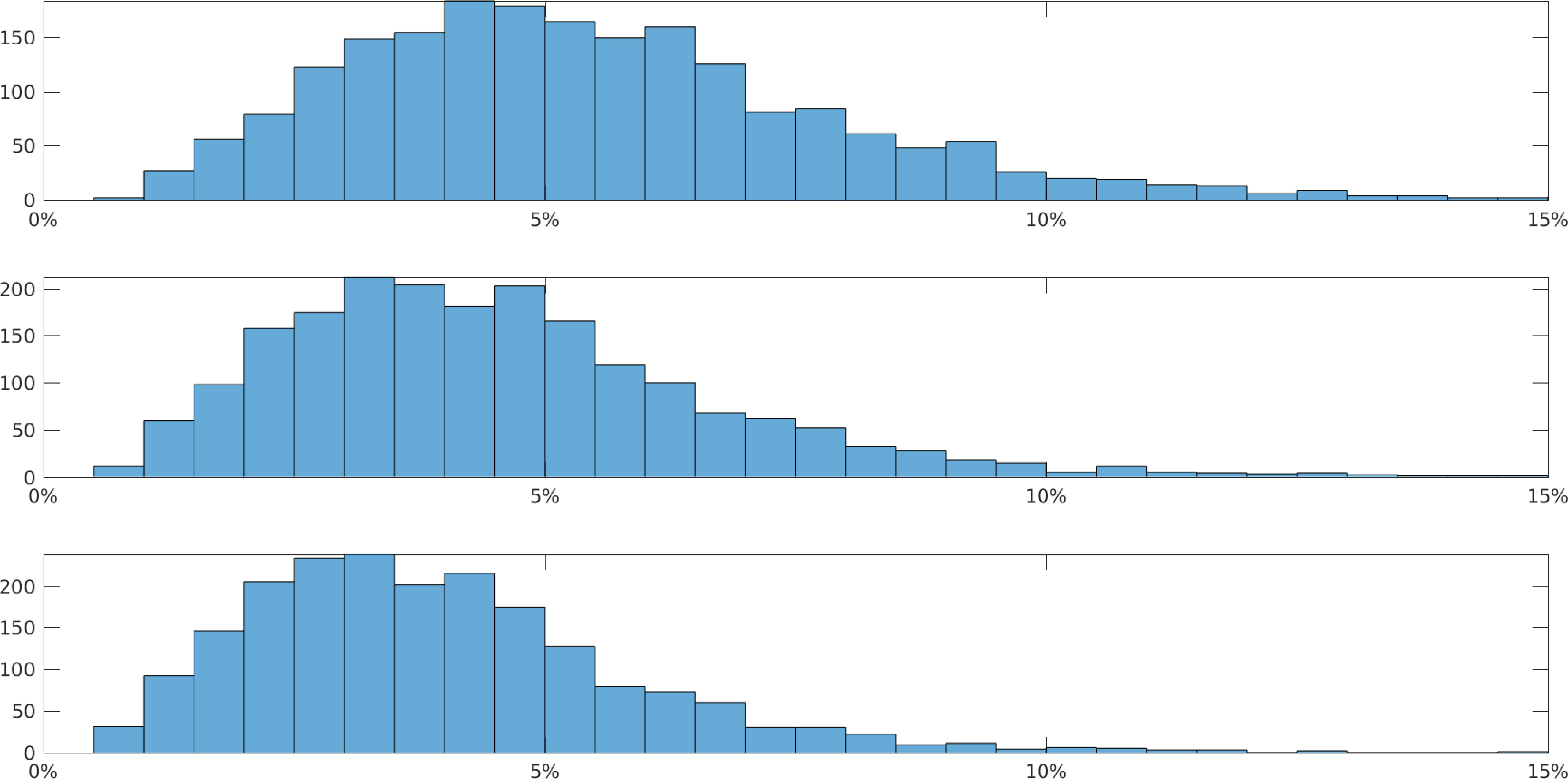}
\end{center}
\caption{\label{fig:ERRORS}
Occurrences of errors for the approximation of the dataset with $\kappa = 8, 14, 19$. On the horizontal axis: the error by pixel (\ref{eq:DELTA}). On the vertical axis: the corresponding number of images for the error.
}
\end{figure} 

Figure \ref{fig:lattice} shows the lattice of translates $\Lambda$ and its dual $L$ used for the presented results, while Figure \ref{fig:ERRORS} shows the distribution of the following adimensional uniform rescaling of the quantity minimized in (\ref{eq:error}):
\begin{equation}\label{eq:DELTA}
\Delta = 100*\left(\frac{1}{d^2}\sum_{n \in \Z_d \times \Z_d}|f_i(n) - \Proj_{S(\Phi)}f_i(n)|^2\right)^\frac12/255 = 100*\frac{\|f_i - \Proj_{S(\Phi)}f_i\|_d}{255*d}.
\end{equation}
This quantity measures a \% error obtained as the average square difference by pixel of an image $f_i$ in the dataset from its optimal approximation $\Proj_{S(\Phi)}f_i$, divided by size of the admissible pixel range for 8 bit images $\{0,\dots,255\}$.

\begin{figure}[h!]
\begin{center}
\includegraphics[width=0.28\textwidth]{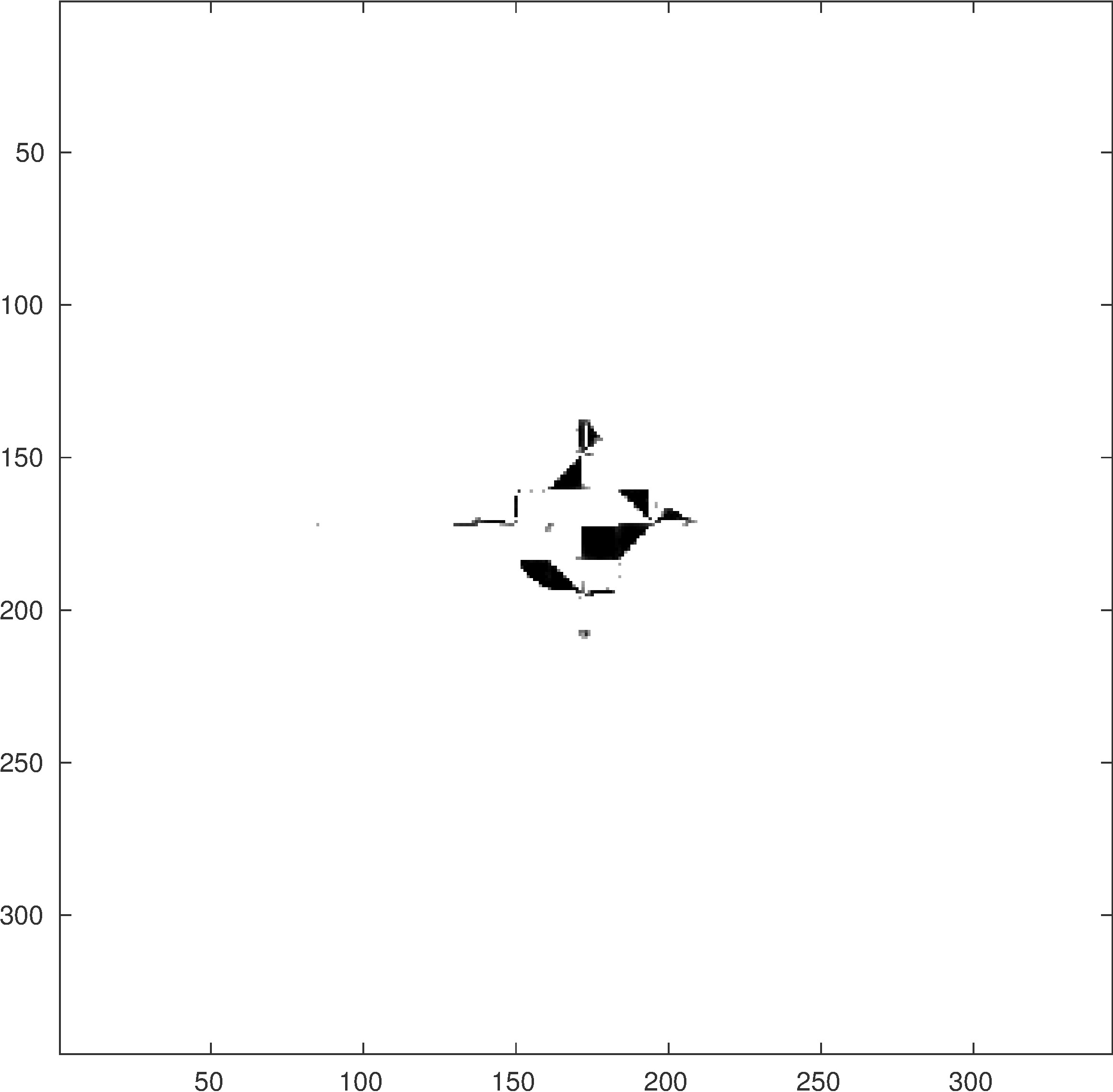} \qquad
\includegraphics[width=0.27\textwidth]{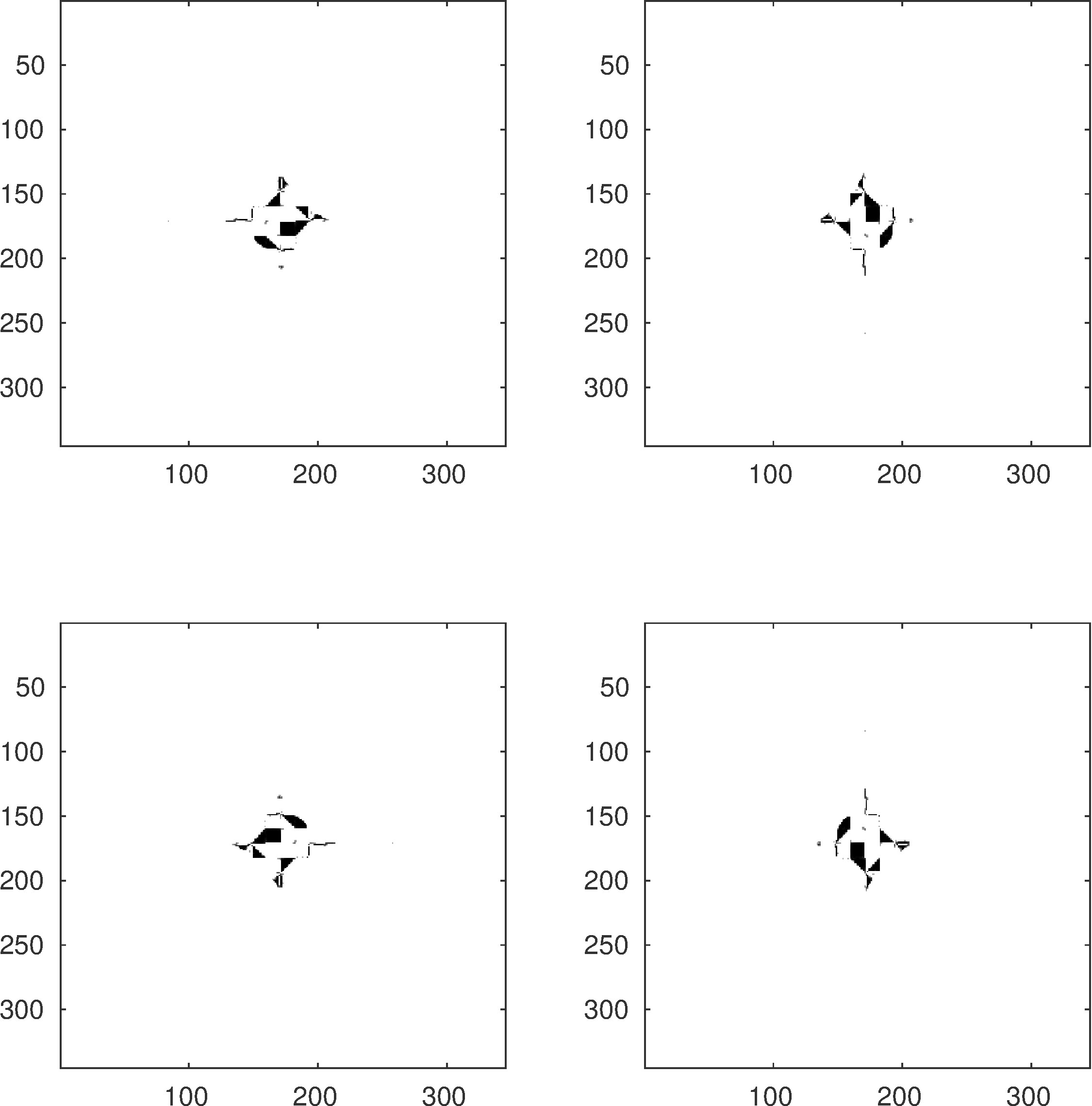} \qquad
\includegraphics[width=0.28\textwidth]{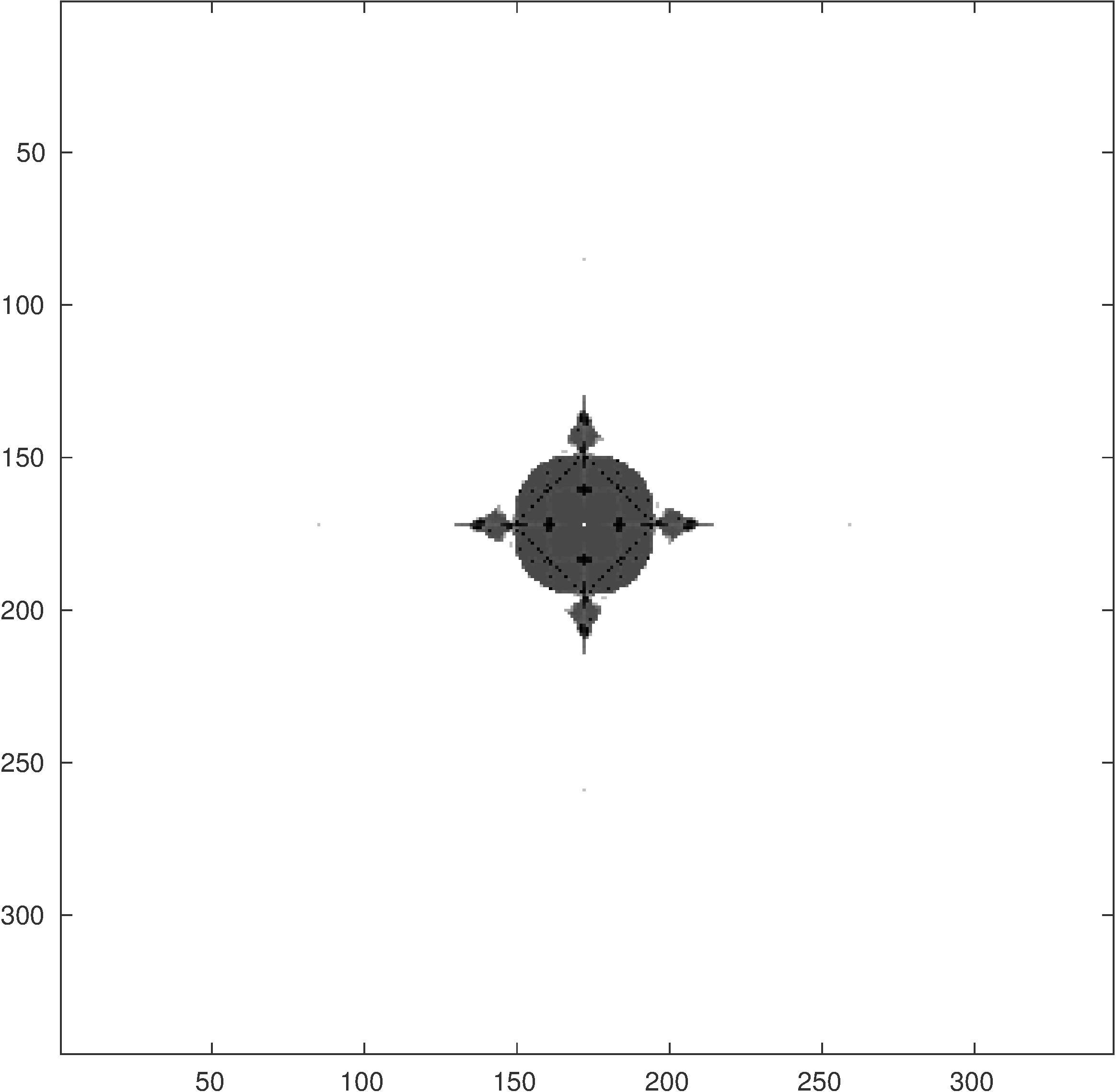}
\end{center}
\caption{\label{fig:FPsi1} 
On the left we show $|\wh{\phi_1}|$, in the center its four rotates, and on the right their sum.
}
\end{figure} 

\begin{figure}[h!]
\begin{center}
\includegraphics[width=0.28\textwidth]{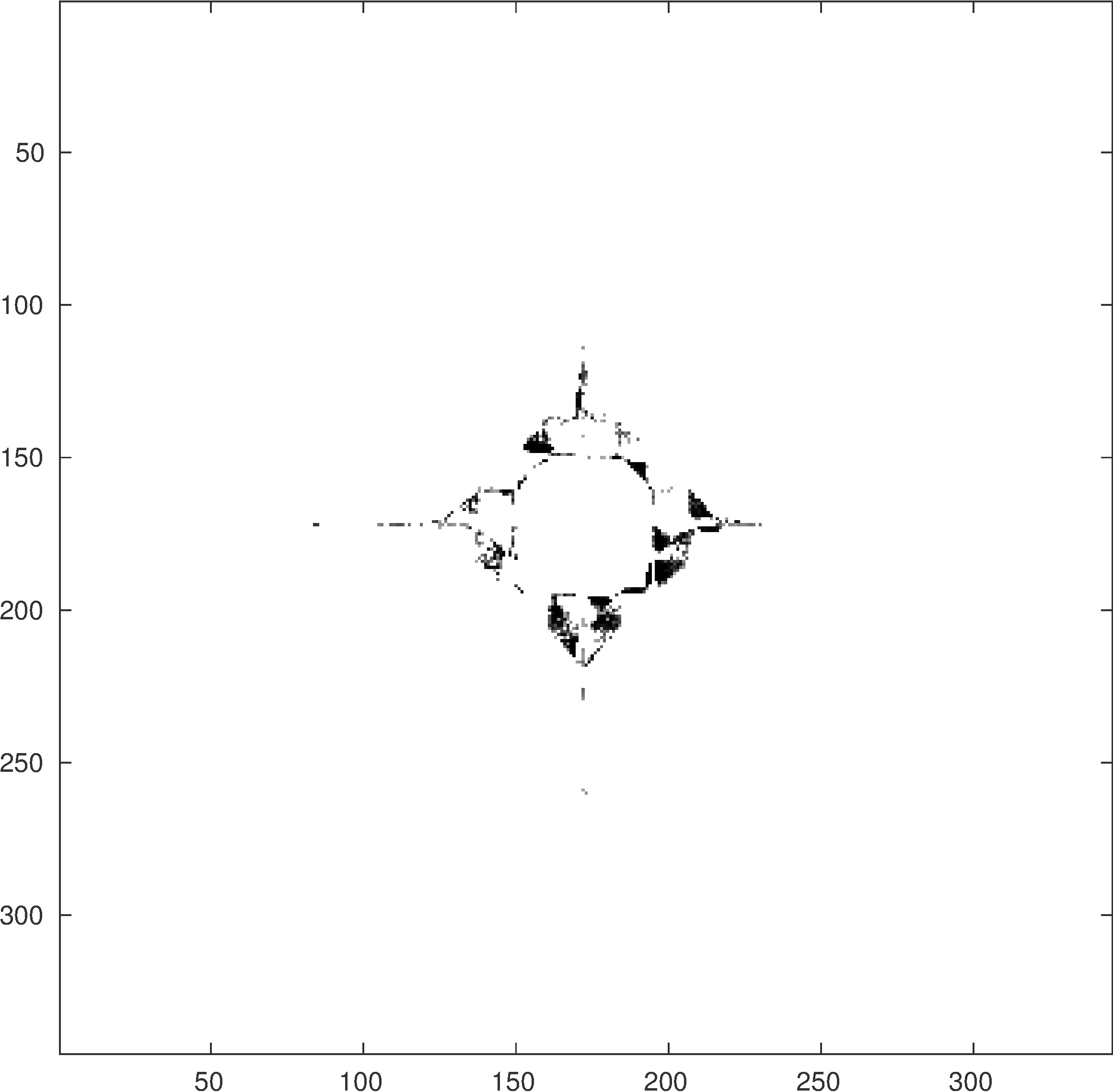} \qquad
\includegraphics[width=0.27\textwidth]{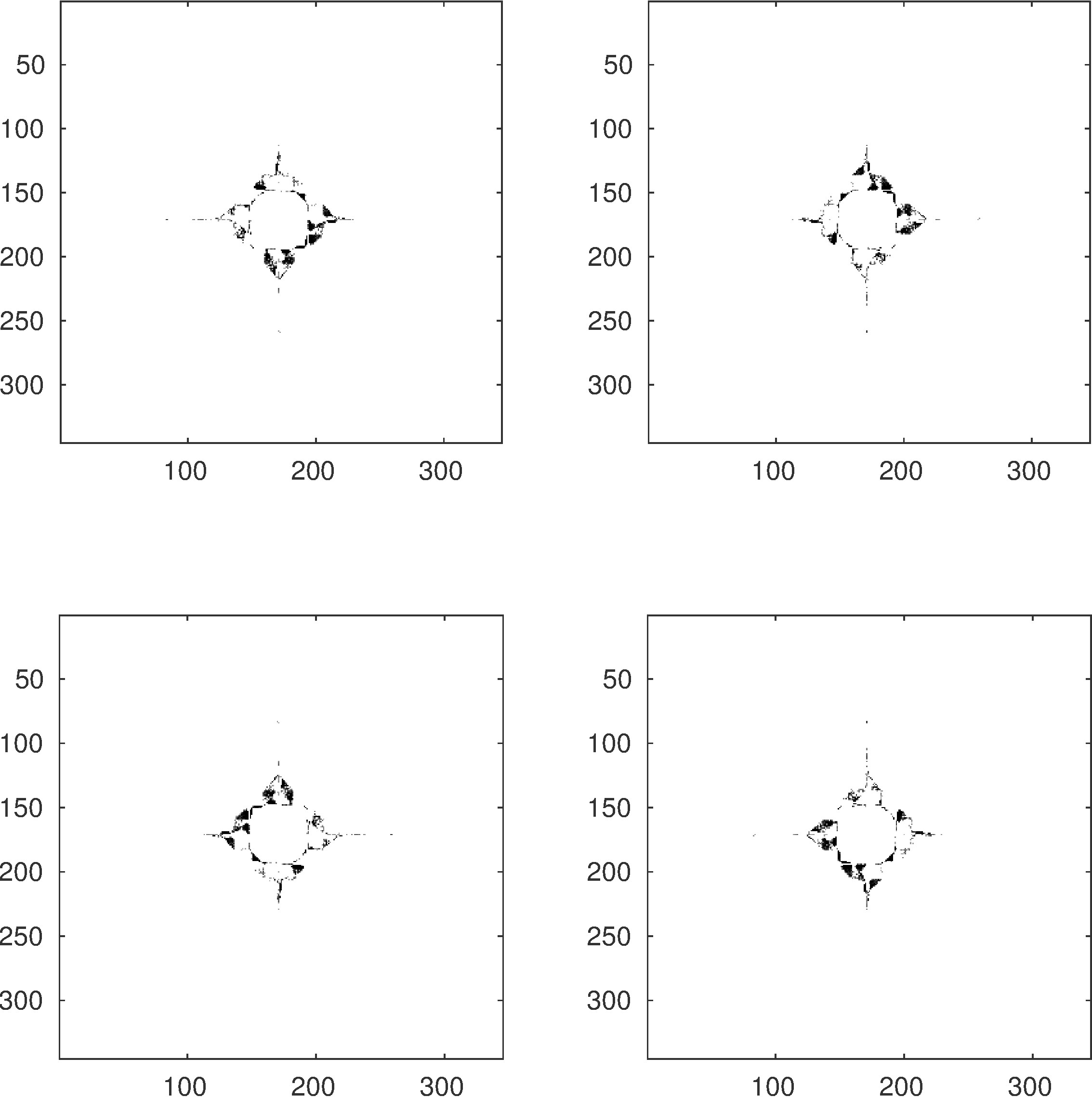} \qquad
\includegraphics[width=0.28\textwidth]{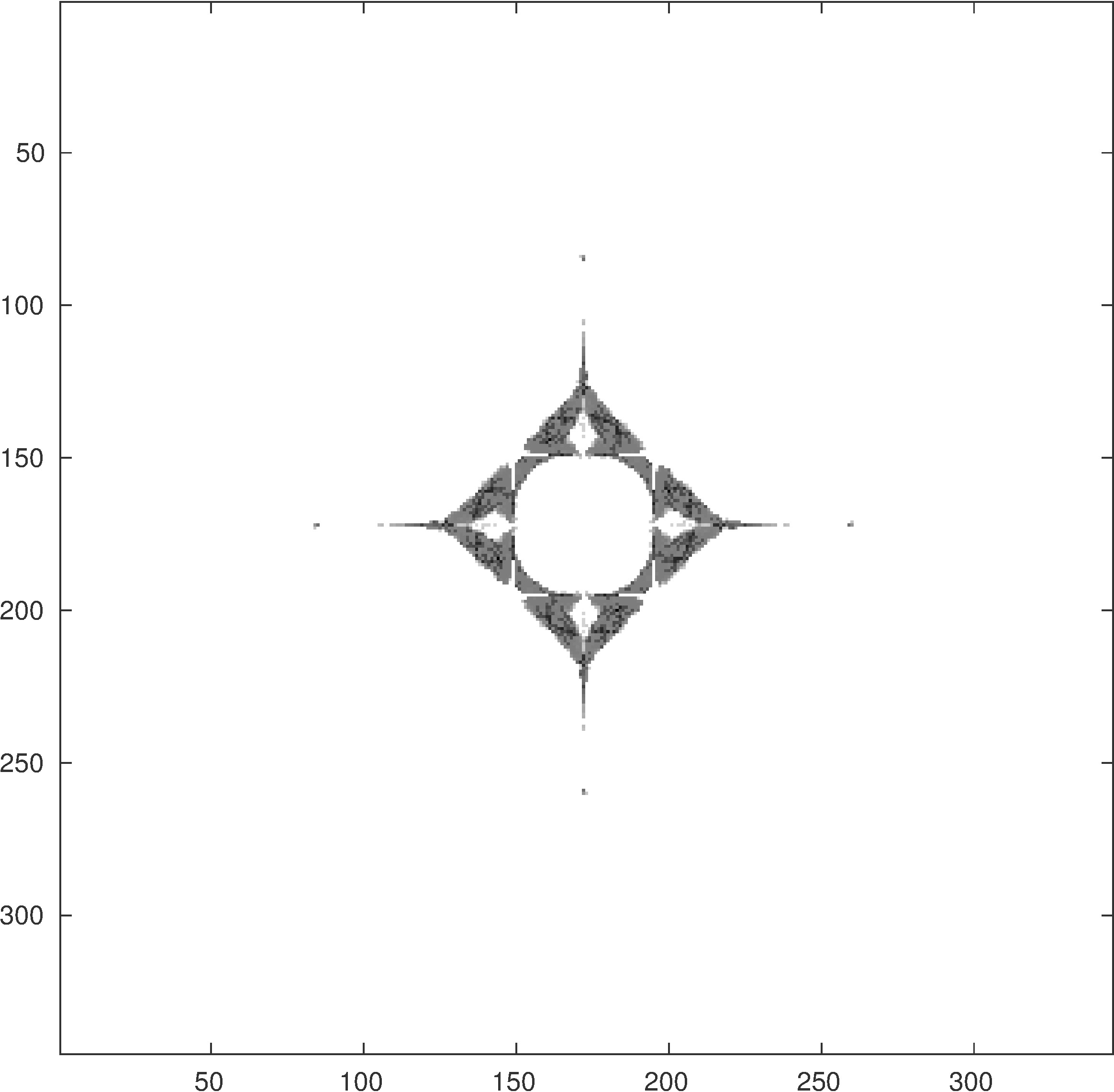}
\end{center}
\caption{\label{fig:FPsi2} 
On the left we show $|\wh{\phi_2}|$, in the center its four rotates, and on the right their sum.
}
\end{figure} 

\begin{figure}[h!]
\begin{center}
\includegraphics[width=0.28\textwidth]{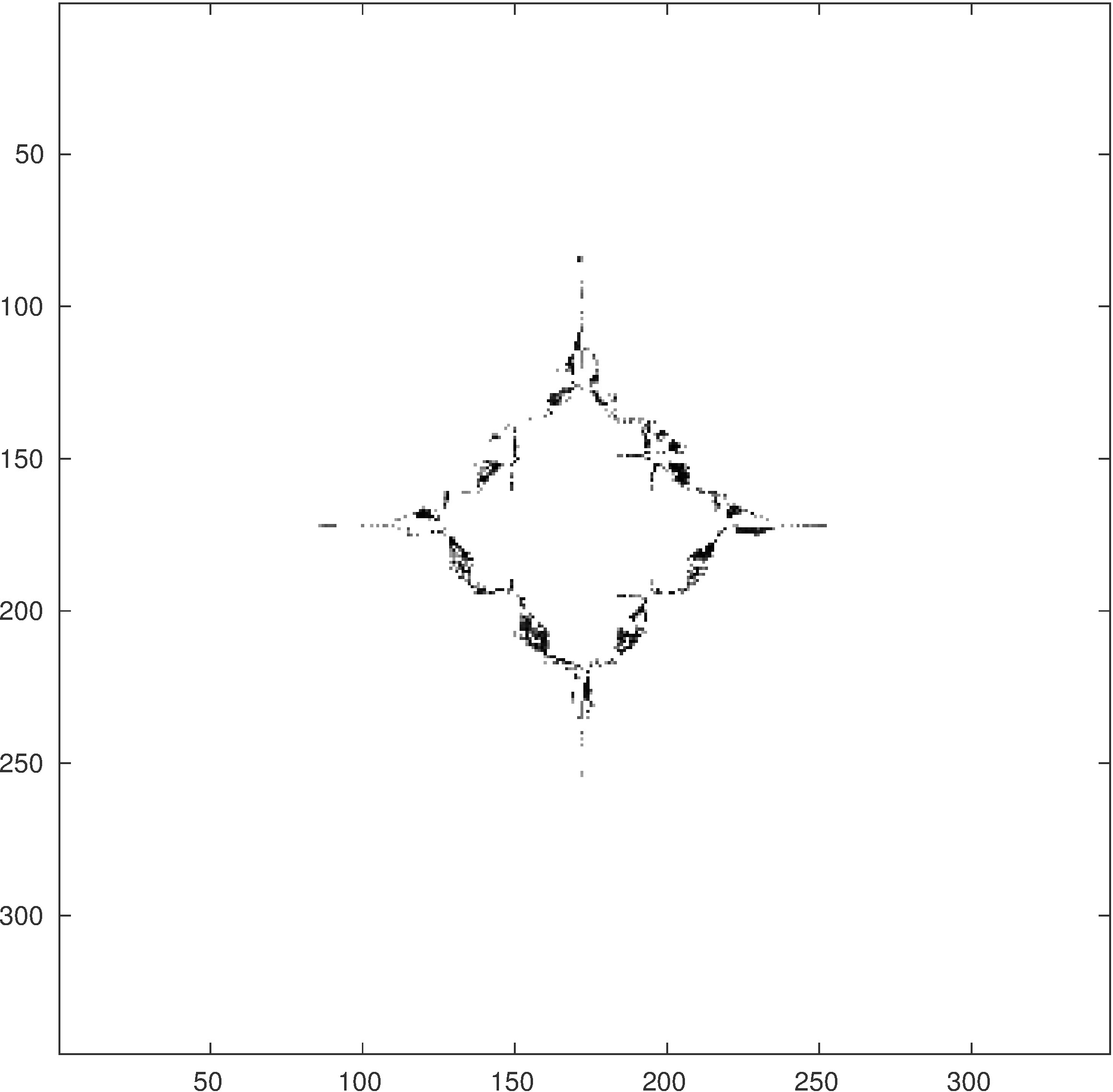} \qquad
\includegraphics[width=0.27\textwidth]{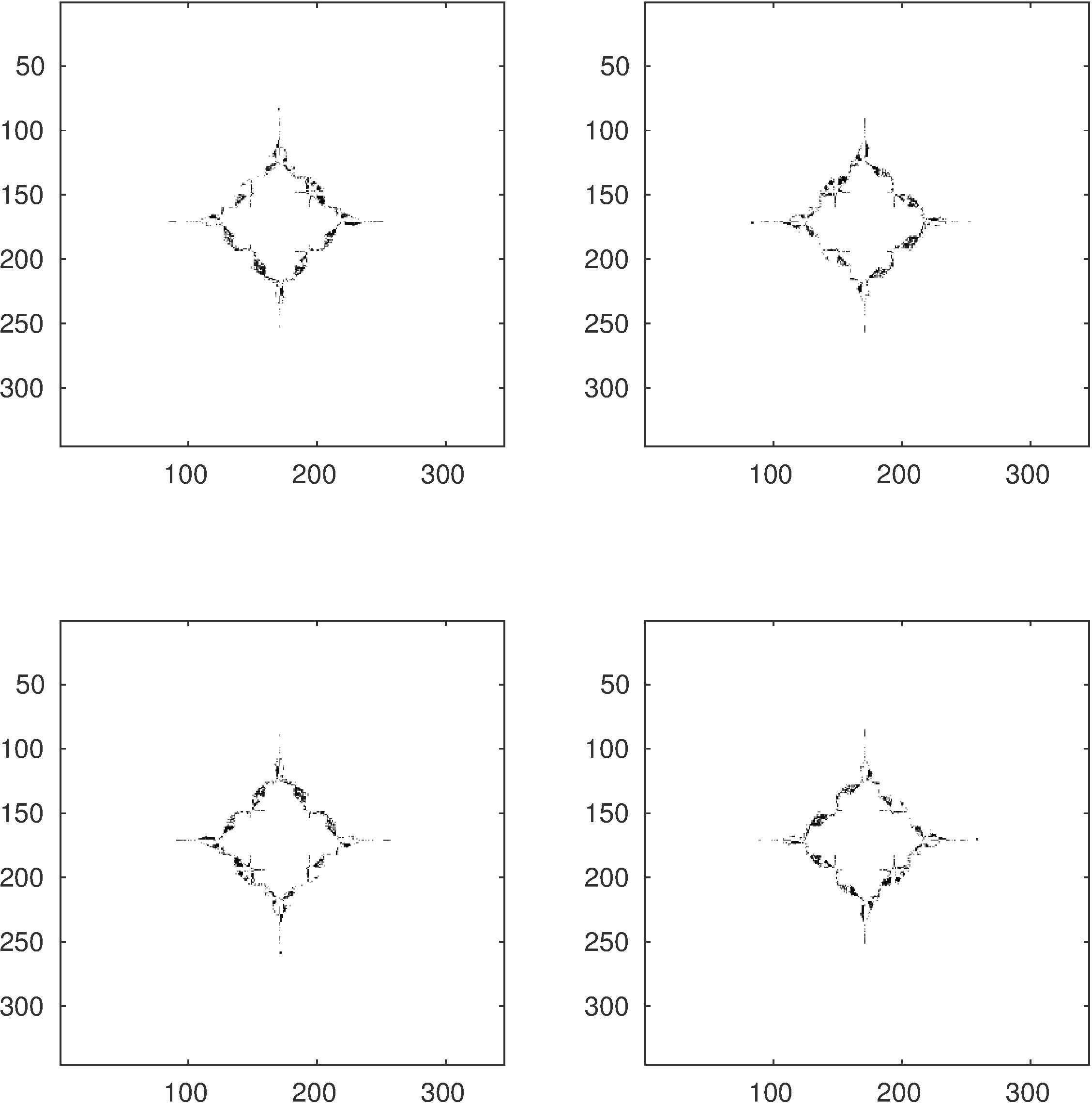} \qquad
\includegraphics[width=0.28\textwidth]{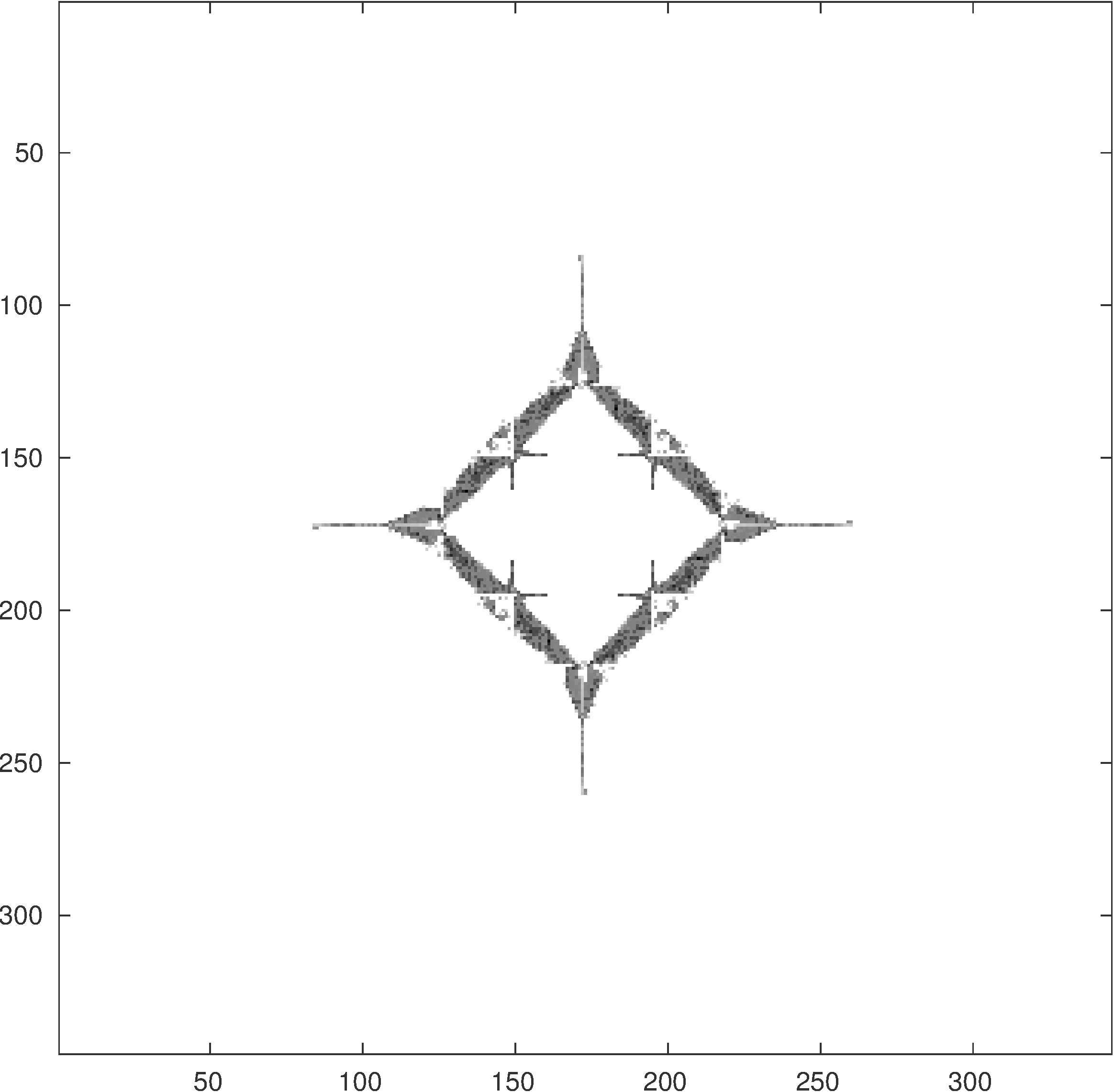}
\end{center}
\caption{\label{fig:FPsi3} 
On the left we show $|\wh{\phi_3}|$, in the center its four rotates, and on the right their sum.
}
\end{figure}

\begin{figure}[h!]
\begin{center}
\includegraphics[width=0.28\textwidth]{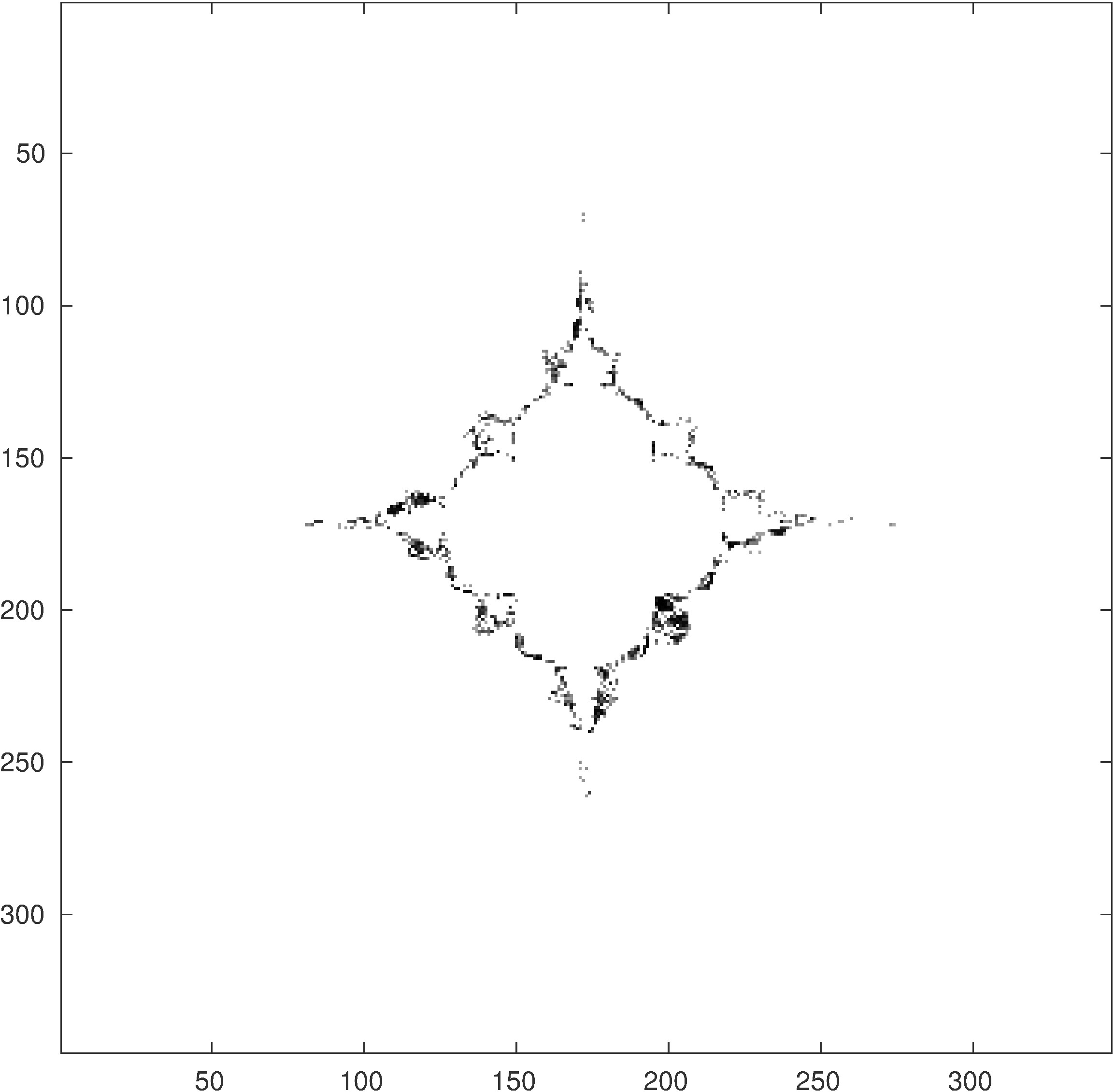} \qquad
\includegraphics[width=0.27\textwidth]{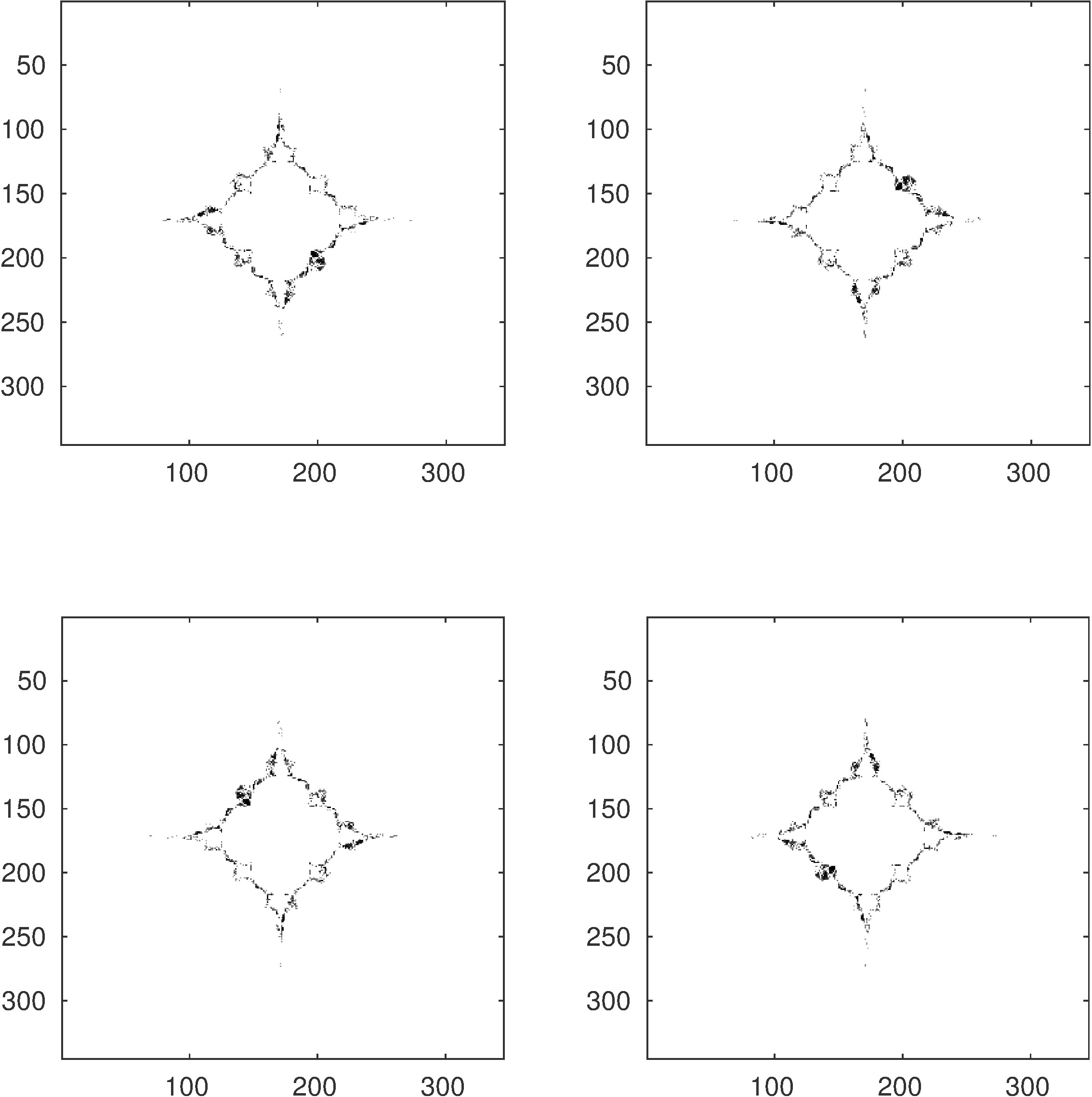} \qquad
\includegraphics[width=0.28\textwidth]{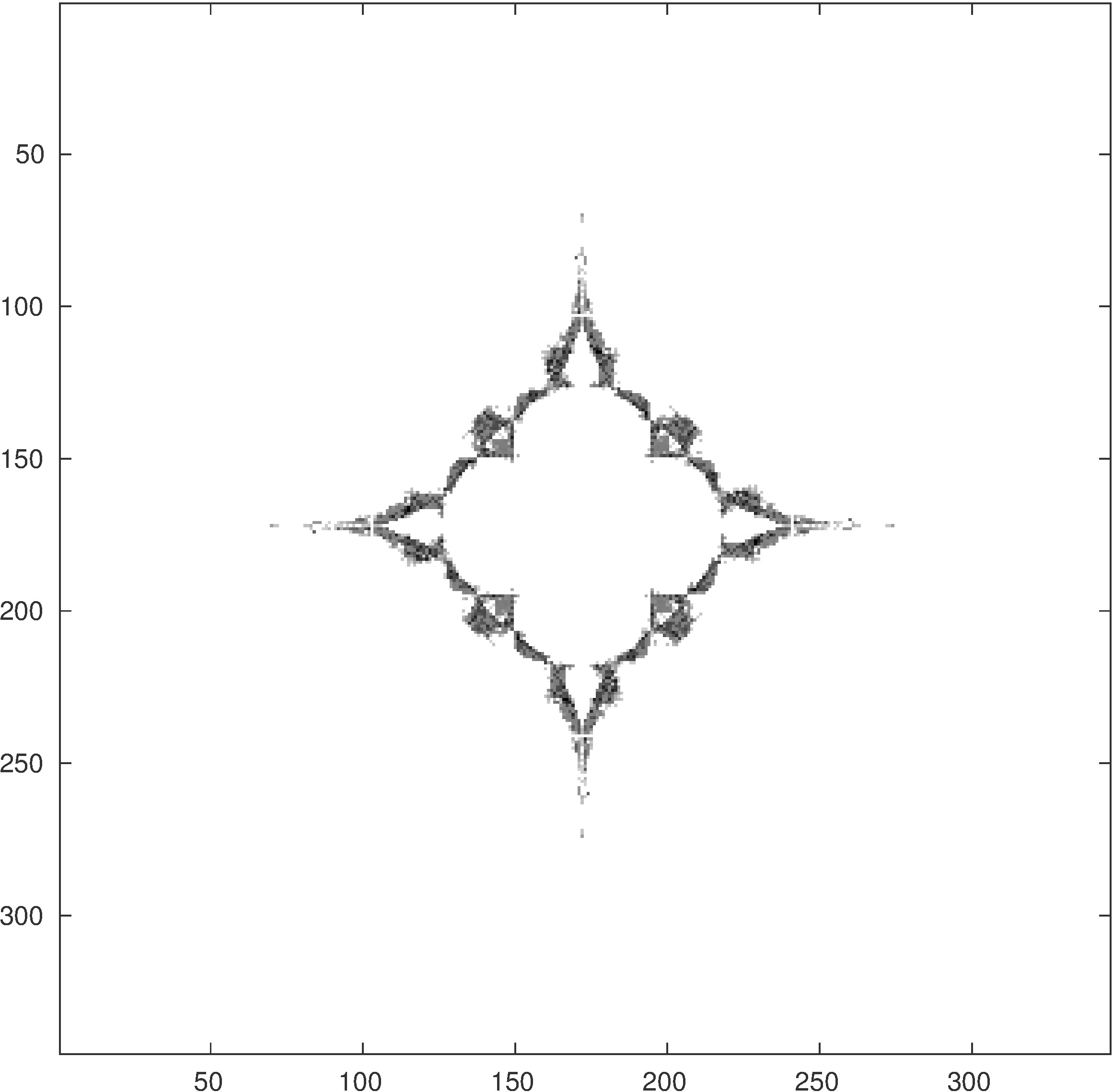}
\end{center}
\caption{\label{fig:FPsi4} 
On the left we show $|\wh{\phi_4}|$, in the center its four rotates, and on the right their sum.
}
\end{figure}

\begin{figure}[h!]
\begin{center}
\includegraphics[width=\textwidth]{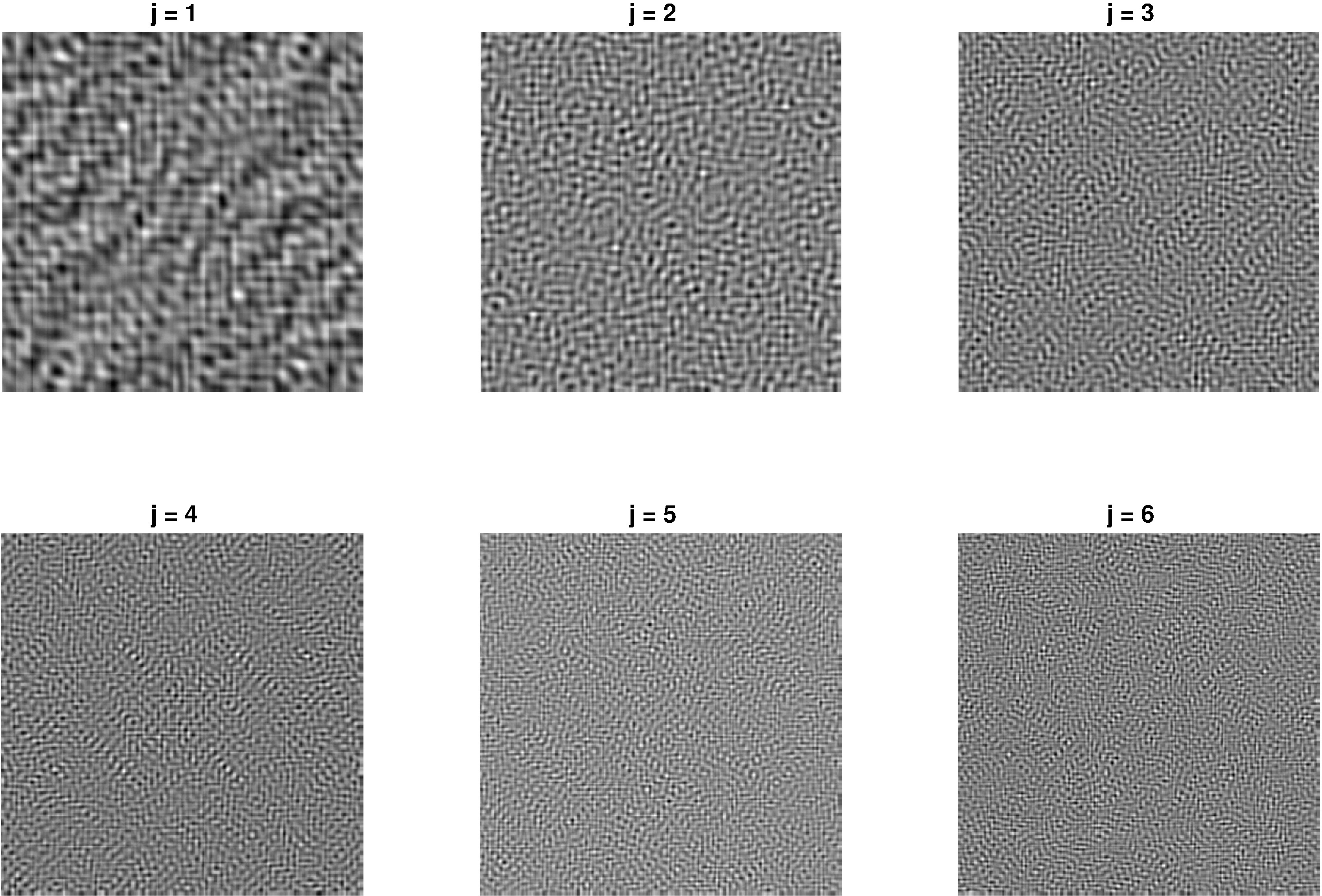}
\end{center}
\caption{\label{fig:GENERATORS}
The real part of the first six optimal generators $\{\phi_j\}_{j = 1}^6$, obtained as in Table 1.
}
\end{figure}

Figures \ref{fig:FPsi1} to \ref{fig:FPsi4} show the absolute value of the Fourier transform of the first four optimal generators $\{\phi_j\}$, obtained as in Table 1. In each figure, in the center we have displayed the four rotates of the corresponding generator, and on the right we have summed them. One can see that the supports of each rotate are almost disjoint, as well as the supports of two different generators. In Figure \ref{fig:GENERATORS} we have shown the real part of the first six generators. One can see that they are delocalized over the whole image area, and are distributed as quasi-periodic patterns of increasing frequency, in accordance with the behavior of their Fourier transform.

\begin{figure}[h!]
\begin{center}
\includegraphics[width=0.3\textwidth]{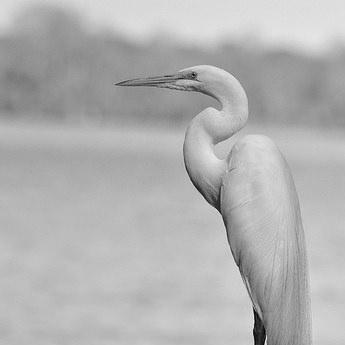} \qquad
\includegraphics[width=0.3\textwidth]{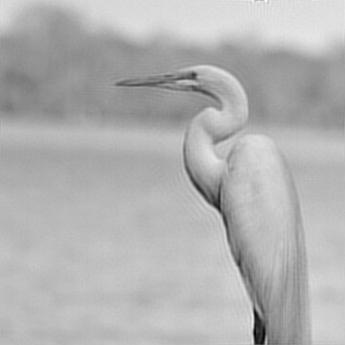} \qquad
\includegraphics[width=0.3\textwidth]{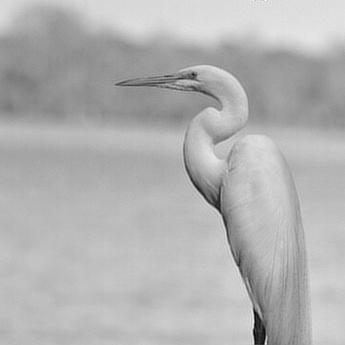}
\end{center}
\caption{\label{fig:693}
Left: original image. Center: $\kappa = 8$, $\Delta = 1.8$\%. Right: $\kappa = 19$, $\Delta = 1.2$\%.
}
\end{figure}

\begin{figure}[h!]
\begin{center}
\includegraphics[width=0.3\textwidth]{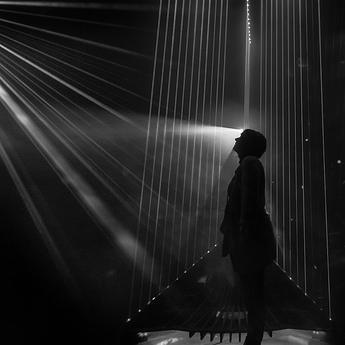} \qquad
\includegraphics[width=0.3\textwidth]{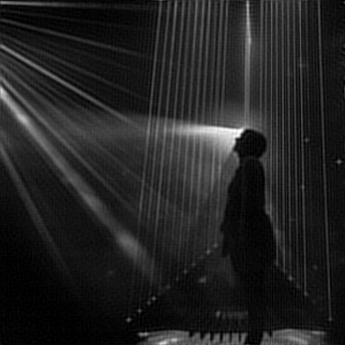} \qquad
\includegraphics[width=0.3\textwidth]{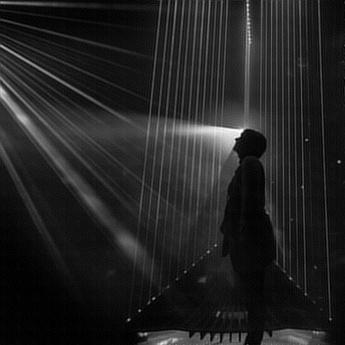}
\end{center}
\caption{\label{fig:329}
Left: original image. Center: $\kappa = 8$, $\Delta = 3$\%. Right: $\kappa = 19$, $\Delta = 1.8$\%.
}
\end{figure}

\begin{figure}[h!]
\begin{center}
\includegraphics[width=0.3\textwidth]{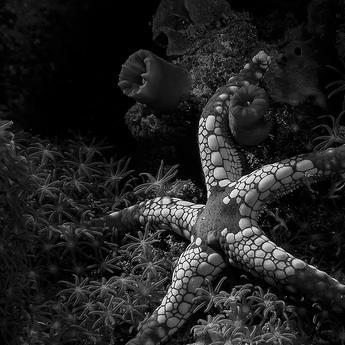} \qquad
\includegraphics[width=0.3\textwidth]{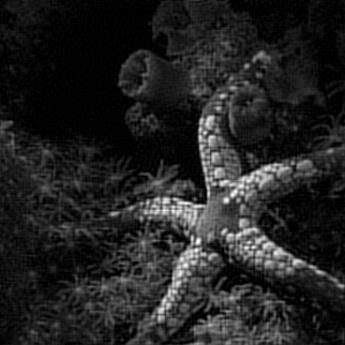} \qquad
\includegraphics[width=0.3\textwidth]{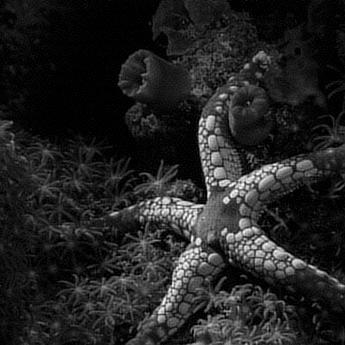}
\end{center}
\caption{\label{fig:661}
Left: original image. Center: $\kappa = 8$, $\Delta = 6.1$\%. Right: $\kappa = 19$, $\Delta = 4.5$\%.
}
\end{figure}

\begin{figure}[h!]
\begin{center}
\includegraphics[width=0.3\textwidth]{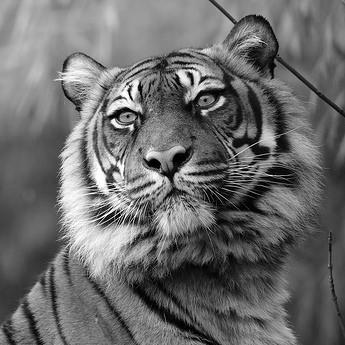} \qquad
\includegraphics[width=0.3\textwidth]{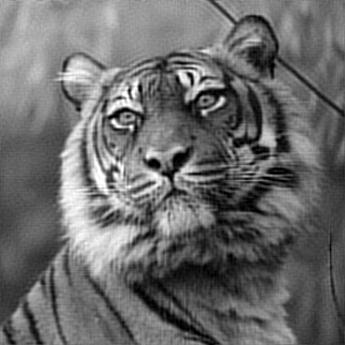} \qquad
\includegraphics[width=0.3\textwidth]{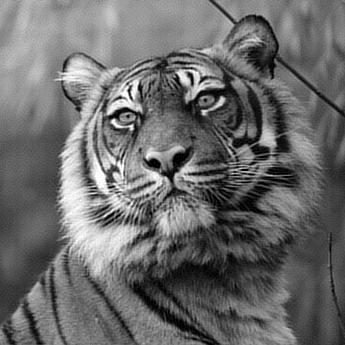}
\end{center}
\caption{\label{fig:1434}
Left: original image. Center: $\kappa = 8$, $\Delta = 6.1$\%. Right: $\kappa = 19$, $\Delta = 4.6$\%.
}
\end{figure}

\begin{figure}[h!]
\begin{center}
\includegraphics[width=0.3\textwidth]{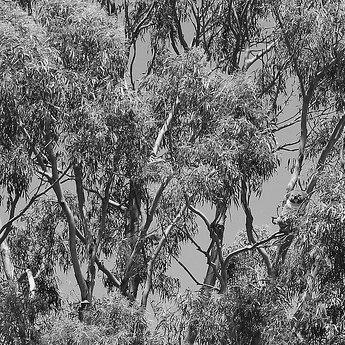} \qquad
\includegraphics[width=0.3\textwidth]{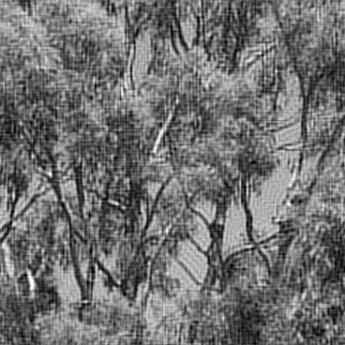} \qquad
\includegraphics[width=0.3\textwidth]{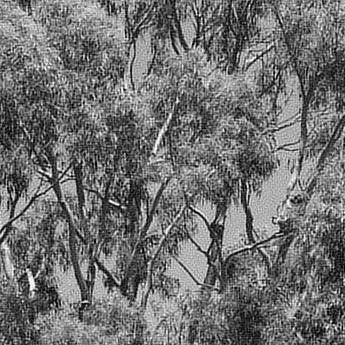}
\end{center}
\caption{\label{fig:471} 
Left: original image. Center: $\kappa = 8$, $\Delta = 16.5$\%. Right: $\kappa = 19$, $\Delta = 12.7$\%.
}
\end{figure}

Figures \ref{fig:693} to \ref{fig:471} show some approximated images, of average behavior in their error range (chosen among the most appealing ones in the dataset) for $\kappa = 8$ and $\kappa = 19$. Their errors (\ref{eq:DELTA}) are given in the captions. It is possible to see that both textures and smooth areas are captured, that the dynamical range is mantained, and that the average pixel error does not always correspond to the visually perceived quality.

\begin{figure}[h!]
\begin{center}
\includegraphics[width=0.3\textwidth]{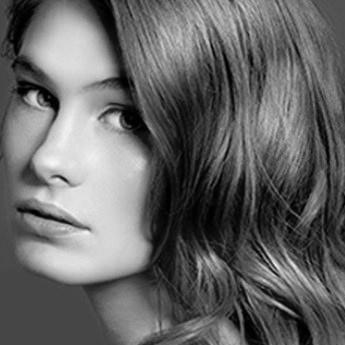} \qquad
\includegraphics[width=0.3\textwidth]{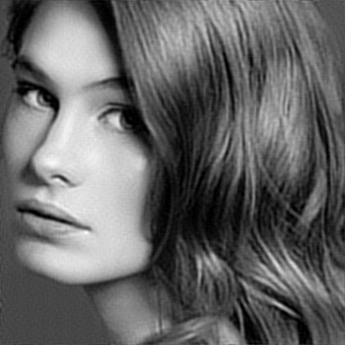} \qquad
\includegraphics[width=0.3\textwidth]{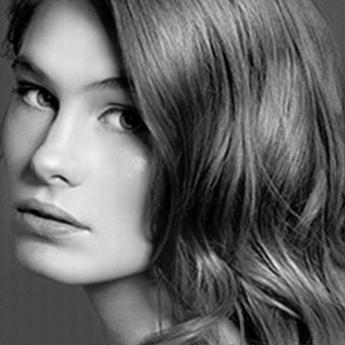}
\end{center}
\caption{\label{fig:898}
Left: original image. Center: $\kappa = 8$, $\Delta = 2.3$\%. Right: $\kappa = 19$, $\Delta = 0.9$\%.
}
\end{figure} 

\acknowledgments
The first author wishes to thank Demetrio Labate for interesting discussions on the present paper and its context.

This project has received funding from the European Union's Horizon 2020 research and innovation programme under the Marie Sk{\l}odowska-Curie grant agreement No 777822. In addidion, D. Barbieri and E. Her\-n\'andez were
supported by Grant MTM2016-76566-P (Ministerio de Ciencia, Innovaci\'on y Universidades, Spain). C. Cabrelli and U. Molter were supported by Grants UBACyT 20020170100430BA (University of Buenos Aires), PIP11220150100355 (CONICET) and PICT 2014-1480 (Secretary of Science and Technology from Argentina).

% References
\bibliography{report} % bibliography data in report.bib
\bibliographystyle{spiebib} % makes bibtex use spiebib.bst

\end{document}